\documentclass[fleqn,usenatbib,useAMS]{mnras}

\usepackage{graphicx}	% Including figure files
\usepackage{amsmath}	% Advanced maths commands
\usepackage{amssymb}	% Extra maths symbols
\usepackage{multicol}        % Multi-column entries in tables
\usepackage{bm}		% Bold maths symbols, including upright Greek
\usepackage{pdflscape}	% Landscape pages
\usepackage{stfloats}

\newcommand{\pa}{\partial}
\newcommand{\mb}{\boldsymbol}

\newcommand{\bgeq}{\begin{equation}}
\newcommand{\edeq}{\end{equation}}

\defcitealias{Bai_2017}{B17}

\usepackage[T1]{fontenc}
\usepackage{ae,aecompl}

\usepackage{newtxtext,newtxmath}
\title[]{Dust Transport in Protoplanetary Disks with Wind-driven Accretion}

\author[Hu \& Bai]{
Zitao Hu$^{1}$
and
Xue-Ning Bai$^{1,2}$%
\thanks{E-mail: \href{mailto:xbai@tsinghua.edu.cn}{xbai@tsinghua.edu.cn}}%
\\
$^{1}$Department of Astronomy,
Tsinghua University, Beijing 100084, China\\
$^{2}$Institute for Advanced Study,
Tsinghua University, Beijing 100084, China
}

\date{Last updated 2021 January 27; in original form 2021 January 27}

\pubyear{2021}

\begin{document}
\label{firstpage}
\pagerange{\pageref{firstpage}--\pageref{lastpage}}
\maketitle

\begin{abstract}
It has recently been shown that the inner region of protoplanetary disks (PPDs) is governed by wind-driven accretion, and the resulting accretion flow showing complex vertical profiles. Such complex flow structures are further enhanced due to the Hall effect, especially when the background magnetic field is aligned with disk rotation. We investigate how such flow structures impact global dust transport via Monte-Carlo simulations, focusing on two scenarios. In the first scenario, the  toroidal magnetic field is maximized in the miplane, leading to accretion and decretion flows above and below.  In the second scenario, the toroidal field changes sign across the midplane, leading to an accretion flow at the disk midplane, with decretion flows above and below. We find that in both cases, the contribution from additional gas flows can still be accurately incorporated into the advection-diffusion framework for vertically-integrated dust transport, with enhanced dust radial ({pseudo-})diffusion up to an effective $\alpha^{\rm eff}\sim10^{-2}$ for strongly coupled dust, even when background turbulence is weak $\alpha<10^{-4}$. Dust radial drift is also modestly enhanced in the second scenario. We provide a general analytical theory that accurately reproduces our simulation results, thus establishing a framework to model global dust transport that realistically incorporates vertical gas flow structures. We also note that the theory is equally applicable to the transport of chemical species.
\end{abstract}


%%%%%%%%%%%%%%%%%%%%%%%%%%%%%%%%%%%%%%%%%%%%%%%%%%

%%%%%%%%%%%%%%%%% BODY OF PAPER %%%%%%%%%%%%%%%%%%

% The MNRAS class isn't designed to include a table of contents, but for this document one is useful.
% I therefore have to do some kludging to make it work without masses of blank space.
%\begingroup
%\let\clearpage\relax
%\tableofcontents
%\endgroup
%\newpage

\section{Introduction}\label{intro}

Dust growth and transport in protoplanetary disks (PPDs) mark the initial stage of planet formation.
Both processes sensitively depend on the gas dynamics of PPDs, particularly on
disk structure and level of turbulence. It is well known that radial pressure gradient in disks leads to
radial drift, typically from outside in, and is more significant as grains grow larger \citep{1977MNRAS.180...57W}.
Disk turbulence leads to dust diffusion, which smears dust spatial distribution towards being well mixed with the
gaseous disk \citep{1993Icar..106..102C, Youdin_2007}. These processes eventually determine the size
distribution as well as the spatial distribution of dust particles, setting the stage for planetesimal formation
\citep{Testi_2014}. The transport processes are also important for understanding the delivery of different types of materials to planetary objects.

There have been multiple lines of evidence on dust transport in PPDs and the solar system.
Radial drift makes the dust disk more compact than the gas disk, which has also been established
from recent observations of nearby PPDs by the Atacama Large Millimeter/sub-millimeter
Array (ALMA) (e.g., \citealt{Ansdell_2018}), and the dust disk is characterized by a sharp outer boundary
\citep{Birnstiel_2013, Testi_2014}. 
The level of turbulence in disks, and hence the strength of dust diffusion, is under debate. Direct measurements
tends to favor weak turbulence (e.g., \citealt{Flaherty_2017,Teague_etal18}, but see \citealt{Flaherty_2020}), which is consistent with
the fact that the dust layer in the HL Tau disk is very thin \citep{Pinte_2015}. On the other hand, modeling of
dust ring width from some ALMA disk observations suggests level of turbulence is strong
\citep{Dullemond_2018,Rosotti_2020}.
In the solar system, the presence of crystalline silicates in comets \citep{1989ApJ...341.1059C,1999ApJ...517.1034W} is widely
interpreted as evidence of large-scale mixing \citep{2002A&A...384.1107B}, particularly the outward
transport of materials processed from the hot inner disk. This conclusion is further supplemented as crystaline sillicates are also widely observed in other protoplanetary disks \citep{vanBoekel2004,Watson2009,Olofsson2009}.
Moreover, dust samples of comet 81P/Wild 2 returned from the Stardust mission, which contain refractory inclusions that formed
in the hot inner disk region \citep{Brownlee_2007}. 
There is also meteoritic evidence based on isotopic
abundances towards large-scale mixing and outward transport (e.g., \citealt{Williams23426}).
Such large-scale mixing generally requires strong turbulence that acts during the early stages of PPD evolution
\citep{Hughes_2010}.

Dust transport in disks is conventionally modeled in the advection-diffusion framework, where advection arises from intrinsic accretion velocity from disk gas, as well as radial drift due to radial pressure gradient, and diffusion arises from disk turbulence. Simple one-dimensional models by itself are difficult to reconcile with evidence of large-scale mixing, as both accretion and radial drift tend to bring dust inward. More complex models consider disk vertical structure. Under the assumption of viscously-driven disk accretion {with constant $\alpha-$ viscosity}, it was found that disks exhibit meridional outflows in the midplane region \citep{Urpin1984,TakeuchiLin02}, which has subsequently been applied to model global dust transport calculations to account for large-scale mixing (e.g., \citealp{KellerGail2004,Ciesla2007,Ciesla2009}). To what extent meridional outflow should exist is debatable {depending on disk properties} (e.g., \citealp{Fromang2011,PhilippovRafikov17}). {For instance, when viscosity is anisotropic dominated by the vertical component, as in the case for turbulence driven by the vertical shear instability (VSI, \citealp{nelson2013linear}), it was shown that gas flows inward in the midplane region and outward in surface layers \citep{StollKley16,Stoll_etal17}.} Overall, understanding the detailed flow properties, especially their vertical structure in PPDs, is crucial to properly model the process of dust transport.

Closely related to flow properties is the processes that drives angular momentum transport in disks. There has been significant developments over the past decade (e.g., \citealt{turner2014transport}). Conventionally, angular momentum transport is understood as entirely due to turbulence (as considered in viscous disk models), while more recently, by better taking into account the non-ideal
magnetohydrodynamic (MHD) effects in weakly ionized gas, it has been found that the inner few AU region of the disks is largely laminar, with accretion mainly driven by magnetized disk winds (e.g., \citealt{Bai_2013}, \citealt{Gressel_2015}). This would suggest much reduced dust transport, apparently in tension with understandings of solar system mixing and some disk observations.

However, wind-driven accretion could induce internal flow structures that advect dust particles and serve as additional means of dust transport. In particular, thanks to the Hall effect, gas dynamics in PPDs depends on the polarity of the large-scale
poloidal magnetic field threading the disk (e.g., \citealt{wardle1999balbus}). When it is aligned with disk rotation, the Hall-shear
instability (HSI, \citealt{kunz2008linear}) strongly amplifies horizontal magnetic fields \citep{lesur2014thanatology,bai2014hall}, creating a
complex gas vertical flow profiles with gas moving both radially inward and outward in different portions of the vertical
column \citep{Bai_2017}. It is conceivable that such velocity profile can enhance radial diffusion of dust: dust particles
undergo vertical oscillations (due to disk turbulence), and hence experience radial velocities in opposite directions.
Such additional transport led by wind-driven accretion is very different from that of viscously-driven accretion, and potentially has a wide range of varieties depending on the detailed disk microphysics. {We note that mixing by lateral, laminar gas flows with varying speeds is also known as ``pseudo-diffusion".} 
It is yet to quantitatively address how {pseudo-diffusion induced by} such complex flow structure influences dust transport. 

In this paper, we use simulation data from two representative radial locations from global non-ideal MHD simulations of
\citet{Bai_2017} (hereafter \citetalias{Bai_2017}) to illustrate the effect of gas flow structure induced by wind-driven accretion on dust
transport. In Section \ref{sec:model}, we describe the disk model and methods of our Monte-Carlo simulations of dust transport, and
the results are presented in Section \ref{sec:result}. We further develop an analytical theory in Section \ref{sec:calculation} to incorporate the role of complex gas flow structures into a verticall-integrated advection-diffusion equation, with more detail in Appendix \ref{sec:appendix}, that fully accounts for the simulation results and is applicable to any gas flow profiles. The results are discussed for potential applications in Section \ref{sec:discussion} before we conclude in Section \ref{sec:summary}. 

\section{Disk Model and Numerical Method}\label{sec:model}

The main purpose of our work is to provide an effective description of dust transport properties given
the vertical profiles of the gas flow. Since such profiles can vary with disk radius, we only seek for a
local approach and characterize the transport coefficients in the most transparent manner. In doing so, we
focus on a local patch of the disk at a fiducial radius $R$ that is characterized by weak turbulence and
wind-driven gas flows (Section \ref{ssec:disk}). On top of a chosen turbulence level and gas flow profile, we inject test particles that
are aerodynamically coupled to the gas and study their transport properties under the local framework (Section \ref{ssec:dust}). The
model and methodology are described below, together with simulation setup and diagnostics discussed in Section \ref{ssec:setup}.

\subsection[]{Model for gas dynamics}\label{ssec:disk}

In local model at fiducial radius $R_0$, as in shearing-sheet \citep{goldreich1965gravitational}, we
take a Cartesian coordinate system centered on that location with $x,y,z$ axis corresponding to the
radial, azimuthal and vertical directions, and $x\equiv R-R_0$.
We here focus on radial dust transport which occurs within
the bulk disk, where it generally suffices to assume an isothermal equation of state and hydrostatic
equilibrium. In this case, the disk vertical density profile given by
\begin{equation}\label{eq:rhog}
    \rho_g(z)=\frac{\Sigma}{\sqrt{2\pi}H_g}\exp\left(-\frac{z^2}{2H_g^2}\right)\ ,
\end{equation}
where $\Sigma$ is the surface density, $H_g=c_s/\Omega$ is gas pressure scale height, with $c_s$
being the isothermal sound speed and $\Omega$ being the Keplerian angular velocity.

\begin{figure*}
    \centering
    \includegraphics[height=8.4cm, width=14.4cm]{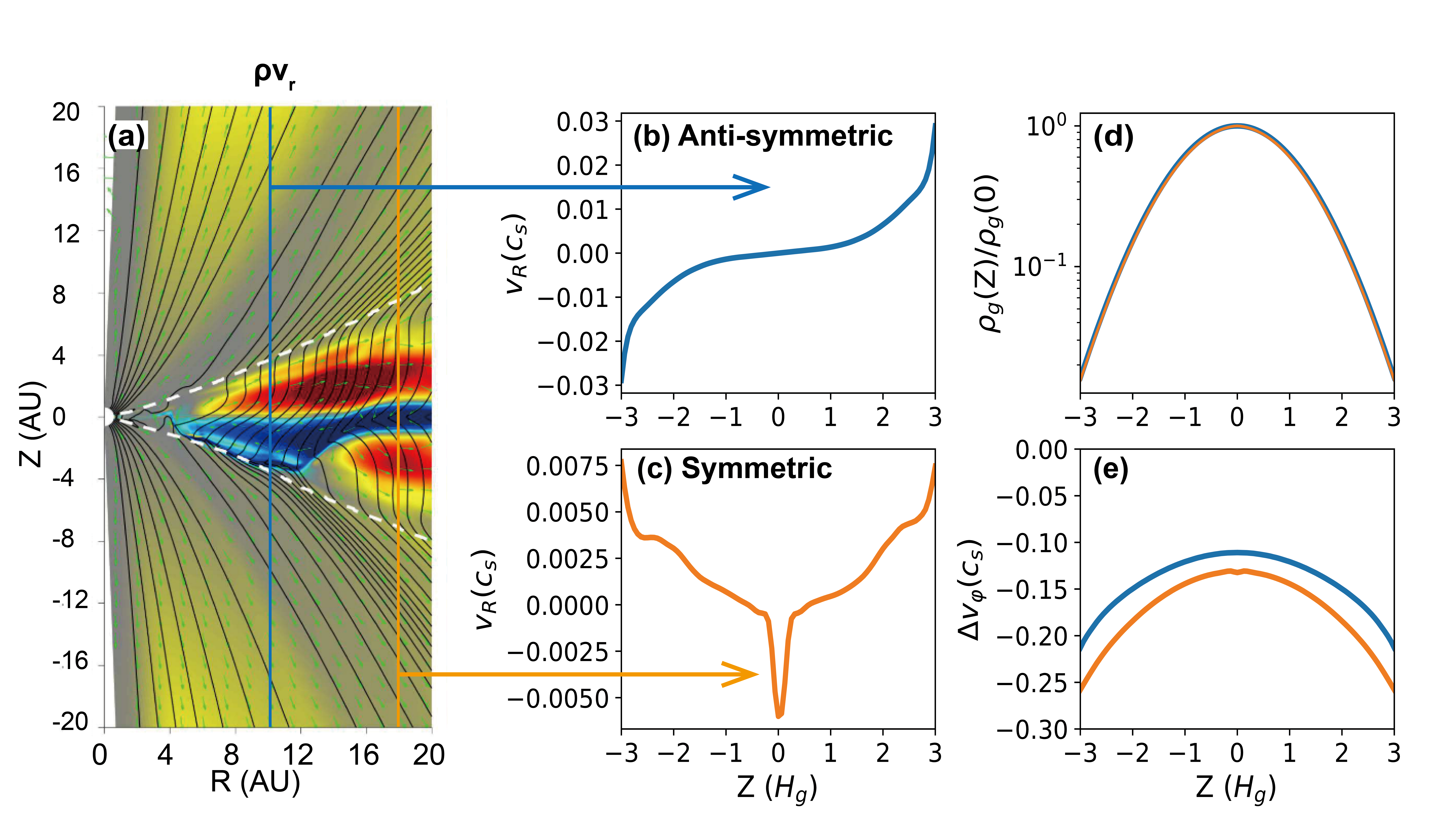}
    \caption{(a) Radial gas flow distribution obtained from MHD simulation, taken from the last panel of
    Figure 7 in \citetalias{Bai_2017}. Panels (b) and (c) are the vertical profiles of radial gas velocity $v_R$ within 3 gas scale height at 10 AU and 18 AU,
    respectively. Panels (d) and (e) show vertical profiles of gas density $\rho_g$ and deviation of azimuthal velocities from Keplerian $\Delta v_\phi$ for both radii.}
    \label{fig:gasflow}
\end{figure*}

The main part of the gas dynamics model is the velocity field. For illustrative purpose, we use the
vertical profile of gas velocity from the fiducial simulation run Fid+ in \citetalias{Bai_2017}, which is the case with net poloidal field aligned with disk rotation. This run is representative in demonstrating
the typical consequences of the HSI, which substantially amplifies horizontal magnetic fields (dominated by the toroidal component). As discussed there, the wind-driven accretion flow structure is determined by the vertical gradient of toroidal field (see Equation (25) of \citetalias{Bai_2017}). Field amplification by the HSI thus enhances the radial gas flow velocity in the disk in both directions. The flow structure resulting from this field configuration is unusual, as shown in Figure \ref{fig:gasflow} (taken partly from Figure 7 of \citetalias{Bai_2017}), and is more complex than the anti-aligned case.\footnote{In the anti-aligned case, there is no strong field amplification to yield complex flow structures, but the flow is not very stable and likely yields weak turbulence (\citealt{2015ApJ...798...84B}, \citealt{simon2015magnetically}, \citetalias{Bai_2017}). More investigations are needed to clarify the nature of the gas flow structure.}
Moreover, it is likely that the solar system formed from a nascent PPD with such aligned field configuration \citep{Weiss2021}. The culprit of this work is to explore the consequences of such complex gas flow structures on dust transport.

For the local model, we choose cylindrical radii $R=10$ AU and $R=18$ AU at the end of run Fid+ of \citetalias{Bai_2017} as two
representative disk radii to carry out our calculations. The density and velocity profiles within $\pm3H_g$ at the
two locations are shown in Figures \ref{fig:gasflow}. Gas density largely follows the
Gaussian profile, and gas velocities remain sub-Keplerian in the bulk disk\footnote{The exact azimuthal velocity
profile could depend on the detailed treatment of thermodynamics, see, e.g., \cite{gressel2020global}, but it does not
affect the calculations nor results of this paper, as it is mainly the radial velocity profile that matters.}. Most notably,
the two radii correspond to two representative configurations of typical flow structures as follows.
\begin{itemize}
\item The ``anti-symmetric" case (10 AU): the vertical profile of radial velocity is characterized by an anti-symmetric
profile, with gas flowing radially inward and outward in either side of the midplane.
 
\item The ``symmetric" case (18 AU): the vertical profile of radial velocity is symmetric about the midplane,
characterized by a midplane inflow (accretion) and radial outflows (decretion) above and below about one
gas scale height.

\end{itemize}
Both anti-symmetric and symmetric velocity profiles are naturally
connected to the magnetic field configurations resulting from the HSI, which were detailed in \citetalias{Bai_2017} and
we only briefly explain below for completeness. 

It is well-known that horizontal field must flip across the disk to achieve a physical wind geometry (so
that there is net accretion flow, e.g., \citealt{Bai_2013}). In the case of the HSI, field amplification is so
strong that in the Hall-dominated inner disk (i.e., 10 AU), the flip occurs only at one (not both!) side of the
disk surface several scale heights above the midplane (as the Hall effect weakens there). As a result,
$B_\phi$ is maximized at the midplane, where its vertical gradient changes sign. As a result, the radial gas
flows in the upper and lower sides of the disk are in opposite directions. The net accretion rate from this
flow configuration is approximately zero due to this anti-symmetry. The net accretion flow is mainly located
at the disk surface where the horizontal field flips, but as it occurs high above the midplane, it hardly affects
dust transport in the bulk disk, and hence is irrelevant for our purpose.

Towards disk outer region (18 AU), the Hall effect weakens, and hence field amplification by the HSI is
not as dramatic. This allows a symmetric system configuration to hold with $B_\phi$ flipping exactly at the
midplane. Correspondingly, the accretion flow resides around the midplane where the flip occurs. On the the
hand, there is still modest field amplification by the HSI, and hence $|B_\phi|$ is maximized near the
midplane (the Hall effect drops with density, and hence the HSI is most effective near the midplane),
creating a vertical gradient of $B_\phi$ towards disk upper layers, which leads to decretion flows there.

The characteristic radial flow velocities from the simulations are in the range of $0.5-2\%$ of sound speed $c_s$ depending on vertical heights. This is smaller than the maximum radial drift velocity for marginally coupled dust, reaching up to $\sim0.1c_s$, i.e., the deviation from Keplerian velocity. For more strongly coupled dust, radial drift velocity can be much smaller (see next subsection), and hence such radial flow velocities can play a dominant role governing dust transport.

While we mostly keep our study dimensionless for illustration purpose, when needed, we also convert the
result to physical units. In doing so, we consider a radial profile of disk surface density $\Sigma(R)$
and temperature $T(R)$ as functions of cylindrical radius as follows
\begin{equation}
\Sigma(R_{AU})=\Sigma_0R_{AU}^{-1}\ ,\quad
T(R_{AU})=T_0R_{AU}^{-1/2}\ ,
\end{equation}
where $R_{AU}$ is radius normalized to AU, and we choose $\Sigma_0=500\text{g cm}^{-1}$ and
$T_0=200\text{K}$.

Besides the laminar flow structure discussed above, we further consider a weak and turbulent flow component.
We do not distinguish the exact origin of the weak turbulence, which is presumably of driven by
hydrodynamic instabilities (e.g., \citealt{lyra2019initial} and references therein), and this is likely present to keep
small (sub-micron to micron sized) dust suspended to be compatible with disk observations (e.g., \citealt{d2001accretion}). Such turbulence is expected to be weak, characterized by Shakura-Sunyaev $\alpha$ parameter of
the order $\alpha\lesssim10^{-3}$. Assuming isotropic turbulence (which can be relaxed), we can write a gas
diffusion coefficient as
\bgeq
D_g=\alpha c_sH_g\ .
\edeq
We will treat the turbulent component stochastically that is fully characterized by $D_g$. It provides a
baseline for dust particles to exercise random motions to reach different vertical locations, and hence experience
different radial velocities of the background gas flows (see next subsection).
In this paper, the range of $\alpha$ we consider varies from $10^{-5}$ to $3\times10^{-2}$.

Finally, some justifications are in order as we consider the laminar and turbulent components separately.
Recent simulations have shown that hydrodynamic turbulence can co-exist with magnetized wind (\citealt{cui2020global}, at least in the case of the VSI turbulence), with wind dominating angular
momentum transport. One interesting result from this work is that the mean flow structure remains similar to
the laminar wind case without turbulence, thus justifying our treatment. 
Moreover, our study only requires an input of mean gas velocity profile and turbulence level. Even the mean
gas velocity profile is modified by turbulence, one could simply update the velocity profiles and our main
results remain to hold as long as the mean gas velocity shows vertical variations.

\subsection{Monte-Carlo simulations of dust dynamics}\label{ssec:dust}

Dust particles are considered as test particles (no backreaction) subject to the drag force $\vec{F}_d$, given by
\begin{equation}
    \vec{F}_d=-\frac{m_s}{t_s}(\vec{v}_d-\vec{v}_g)\ ,
\end{equation}
where $\vec{v}_g$ and $\vec{v}_d$ are the velocities of gas and dust, respectively, $m_s$ is dust particle
mass, and $t_s$ is the dust stopping time. Without loss of generality, we can interpret $\vec{v}_d$ and
$\vec{v}_g$ as dust and gas velocities relative to local Keplerian velocity, and gas velocity refers to the
velocity of the mean flow (laminar component). 

As usual, we characterize the stopping time by the dimensionless Stokes number
\begin{equation}
{\rm St}=\Omega t_s\ ,
\end{equation}
which reflects how well dust is coupled to disk gas. In this paper, we refer to ${\rm St}$ measured at
the disk midplane, and we consider ${\rm St}$ ranging from $3\times10^{-5}$ to $0.1$. Applying the numbers
to our disk model, they correspond to grain size from $a=6\mu$m to $2$cm at 10 AU and from $a=3.3\mu$m
to $1.1$cm at 18 AU for grains with bulk density $\rho_s=1$g cm$^{-3}$, both are under the Epstein
drag law \citep{epstein1924resistance}, where $t_s=\rho_da/(\rho_gc_s)$.

Without turbulence, motion of individual dust particles follows
\begin{equation}\label{dynamiceq}
 \frac{d\vec{v}_d}{dt}=2v_{d,y}\Omega\hat{x}-\frac{1}{2}v_{d,x}\Omega\hat{y}-\Omega^2z\hat{z}-\frac1{t_s}\left(\mb{v}_d-\mb{v}_g\right)\ ,
\end{equation}
which gives the terminal (equilibrium) velocity as
\begin{equation}\label{vdr}
   \bar{v}_{d,x}=\frac{v_{g,x}}{1+St^2}+\frac{2v_{g,y}}{St+St^{-1}}\ ,
\end{equation}
\begin{equation}\label{vdy}
   \bar{v}_{d,y}=\frac{v_{g,y}}{1+St^2}-\frac{v_{g,x}}{2(St+St^{-1})}\ ,
\end{equation}
\begin{equation}\label{vdz}
    \bar{v}_{d,z}=v_{g,z}-\Omega^2 t_sz\ .
\end{equation}
In our calculations, we take $v_{g,x}$ and $v_{g,y}$ from run Fid+ of \citetalias{Bai_2017}, but assume
$v_{g,z}=0$. We note that despite that $v_{g,z}$ increases towards disk surface accompanied
by wind launching, the wind speed is small (less than $0.005c_s$ within $z=\pm3H$)
and given level of turbulence that we consider here, grains with ${\rm St}\gtrsim10^{-4}$
rarely get lifted to beyond $z=\pm3H_g$ (which requires $\alpha\gtrsim St[\rho_g(0)/\rho_g(3H)]\sim100St$). In situations where this does happen,
these particles are discarded, and we have verified they only constitute of a tiny fraction of particles.

With turbulence, particles receive additional stochastic velocity kicks. In Monte-Carlo simulations,
such velocity kicks leads to dust diffusion, with diffusion coefficient depending on turbulence
properties. When assuming turbulence is isotropic with eddy turnover time being $\sim\Omega^{-1}$, the radial
diffusion coefficient is given by \citep{Youdin_2007}
\begin{equation}
    D_{d}=\frac{1+4St^2}{(1+St^2)^2}D_g\approx\frac{D_g}{1+St^2}\ .
\end{equation}
Given the uncertainties in turbulence properties, we use the latter approximate expression.
While the analogous expression is not available for vertical diffusion coefficient, it generally
suffices to use the same expression, which can be derived following \citet{Youdin_2007} without
considering rotation.

Dust diffusion homogenizes dust concentration ($\rho_d/\rho_g$), giving a diffusive flux
\bgeq\label{eq:Fd}
\vec{F}_d=-\rho_gD_d\nabla\bigg(\frac{\rho_d}{\rho_g}\bigg)=
-D_d\nabla\rho_d+\rho_d\langle\delta\vec{v}\rangle\ ,
\edeq
which can be cast into a normal diffusion together with an additional mean
velocity given by
\bgeq
\langle\delta\vec{v}\rangle\equiv\frac{D_d}{\rho_g}\nabla\rho_g\ ,
\edeq
which corrects for the excess dust diffusion flux from concentration gradient
\citep{charnoz2011three}.

In our Monte-Carlo simulation of dust particles, we update particle positions over a constant
timestep $\Delta t=0.1\Omega^{-1}$. This is typically larger than dust stopping time we consider.
To speed up the calculation, we use the terminal velocity approximation 
(e.g., \citealt{jacquet2011linear, price2015fast}, applicable when $\Delta t\gtrsim t_s$), 
and the particle integration proceeds as \citep{Ciesla2010,charnoz2011three}
\bgeq
\vec{r}(t+\Delta t)=\vec{r}(t)+\vec{v}'_d\Delta t+\sqrt{2D_d\Delta t}N(0,1)\ ,
\edeq
where
\bgeq\label{eq:vd}
\vec{v}'_d=\bar{\vec{v}}_d+\frac{D_d}{\rho_g}\nabla\rho_g
\edeq
is the terminal velocity supplemented by $\langle\delta\vec{v}\rangle$ to correct for
concentration diffusion, and $N(0,1)$ is a random number satisfying the standard normal distribution.
In this approach, only the $\hat{x}$ (or $R$) and $\hat{z}$ components are needed for our study.

We have verified that our methodology and choice of timestep are sufficient to accurately simulate
dust radial drift and diffusion.

\begin{figure*}
    \centering
    \includegraphics[height=15cm, width=15cm]{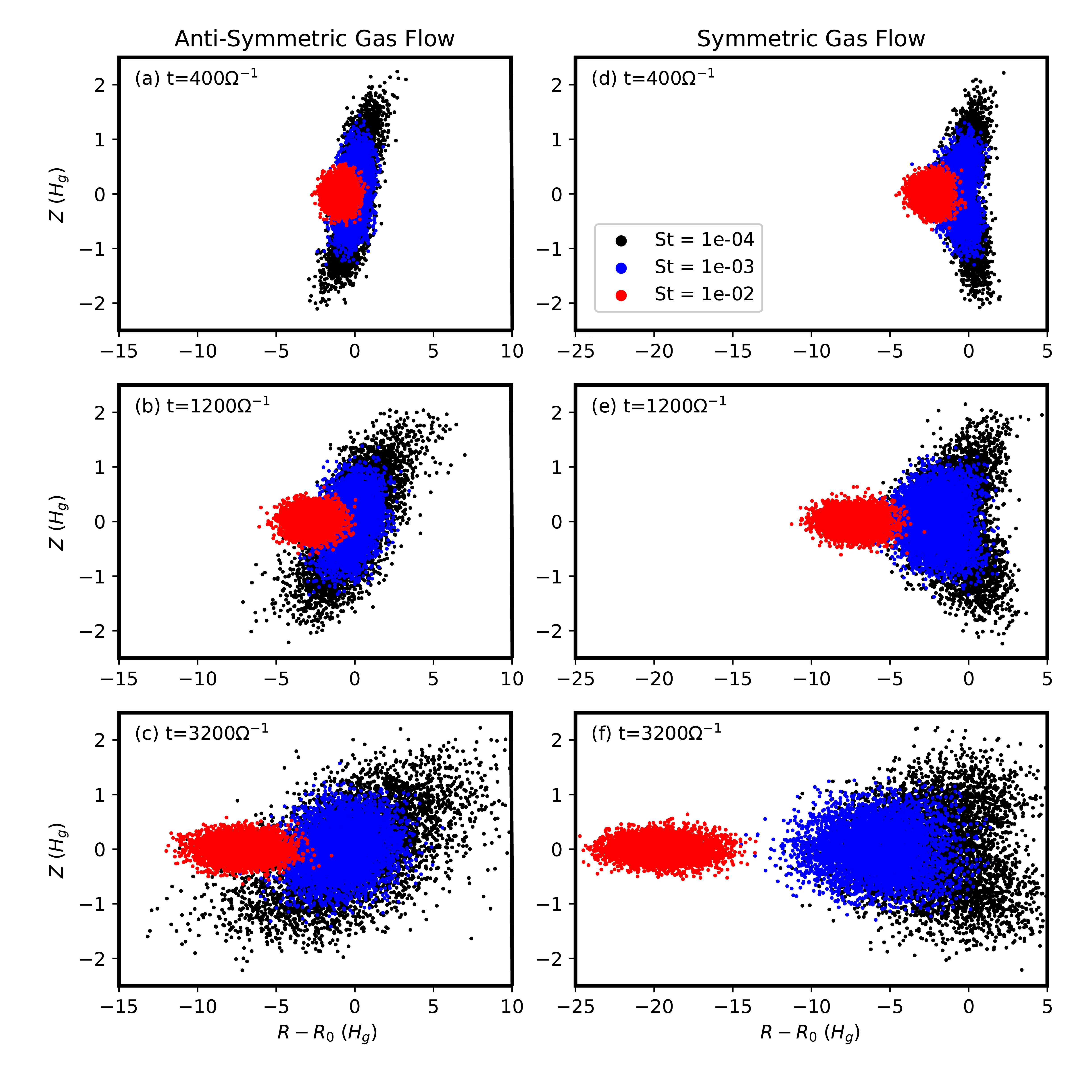}  
    \caption{
    Evolution of dust spatial distribution relative to initial radial location $R_0$, shown at different times
    (from top to bottom), for the antisymmetric (10 AU, left) and symmetric (18 AU, right) gas flow cases
    with $\alpha = 3\times10^{-4}$. Shown are results from a random sub-sample of 5000 dust particles
    for each $St$, indicated by different colors (see legend).}
    \label{fig:scatter}
\end{figure*}

\subsection{Simulation Setup and Diagnostics}\label{ssec:setup}

Our simulation parameters include a choice of gas
velocity profiles (anti-symmetric or symmetric), turbulent diffusivity characterized by $\alpha$
($10^{-5}$ to $0.03$), and dust midplane Stokes number ${\rm St}$ ($3\times10^{-5}$ to $0.1$). 
Given the parameters, in each simulation, we inject about
$2\times10^5$ particles at $t=0$ with spatial coordinates $(x,z)=(0, z)$ and study their
radial transport.  We note that in our local approach, dust motion in the vertical and radial directions
are largely decoupled. Therefore, we set the initial vertical coordinates $z$ of the particles
to satisfy a distribution function that corresponds to the equilibrium particle distribution function.

For gas density $\rho_g(z)$ in hydrostatic equilibrium (Equation \ref{eq:rhog}),
the vertical profile of dust density can be calculated analytically by balancing settling flux
$\rho_d\bar{\vec{v}}_d$ and diffusive flux of Equation (\ref{eq:Fd}) in the vertical direction, which yields
\begin{equation}\label{eq:rhod}
    \rho_d(z)=\rho_{d0}\exp\left(-\frac{z^2}{2H_g^2}\right)\exp\left(\int_0^z\frac{\bar{v}_{d,z}}{D_d}dz\right)\ ,
\end{equation}
where $\rho_{d0}$ is the dust density in the middle plane. We have verified that when turning off
particle motion in the $x,y$ directions, this dust density profile can be accurately maintained in our
Monte-Carlo simulations.

We simulate particle motion for up to $8\times10^4\Omega^{-1}$ and examine the spatial distribution of particles
at constant time intervals. We will discuss the full statistical distribution
from the simulations, and in particular characterize this distribution by computing the vertically-integrated mean speed of radial drift $v_{d,R}^{\textit{eff}}$
and an effective diffusion coefficient $D_{d,R}^{\textit{eff}}\equiv\alpha_{d,R}^{\textit{eff}}c_sH_g$, given by 
\bgeq\label{eq:r2}
    \langle [R(t)-R_0]^2\rangle = (v_{d,R}^{\textit{eff}}t)^2+2D_{d,R}^{\textit{eff}}t\ .
\edeq 
Both of these quantities are likely modified by the contribution from the gas flows
as a result of wind-driven accretion. {In particular, we also call $D_{d,R}^{\textit{eff}}$ {\it pseudo-diffusion} coefficient, which reflects the modification by laminar gas flows.}

For clarity in notation, we use $D_g$ and $D_d$ for intrinsic diffusion coefficient for gas and dust due to background gas turbulence. While we assume they are isotropic in the simulations, anisotropic diffusion can also be considered {(e.g., \citealp{Stoll_etal17})}, where we denote as, e.g., $D_{d,R}$ and $D_{d,z}$ for normal dust diffusion coefficients in radial and vertical directions, used in Section \ref{sec:calculation}. Note that they can also be a function of vertical coordinates. Presence of radial gas flows contributes to additional {(pseudo-)}diffusion, and we write
\bgeq\label{eq:Dtot}
D_{d,R}^{\textit{eff}}\equiv \overline{D_{d,R}}+\Delta D_{d,R}^{\textit{eff}}\ ,
\edeq
where $\Delta D_{d,R}^{\textit{eff}}$ denotes contribution from radial gas flows and
\bgeq\label{eq:DdRbar}
\overline{D_{d,R}}=\int \rho_d(z)D_{d,R}(z)dz\bigg/\int \rho_d(z)dz
\edeq
{is the weighted average of the mean dust radial diffusion coefficient (in similar way as Equation (\ref{eq:vr})).
 We also use the term ``effective" and ``pseudo-" interchangeably in describing vertically-integrated total radial diffusion coefficient.}

Similarly, $v_{g,R}(z)$ denotes vertical velocity profile of the gas. The resulting radial drift velocity profile for dust is given by Equation (\ref{vdr}), and for the remaining of the paper we write it as $v_{d,R}(z)$. We note it has two contributions: one associated with background radial gas flow, and one from background pressure gradient (first and second terms in Equation (\ref{vdr})). Finally, we use $v_{d,R}^{\textit{eff}}$ to denote the mean radial drift speed of the entire particle population.

\begin{figure}
    \centering
    \includegraphics[height=5.6cm, width=8.4cm]{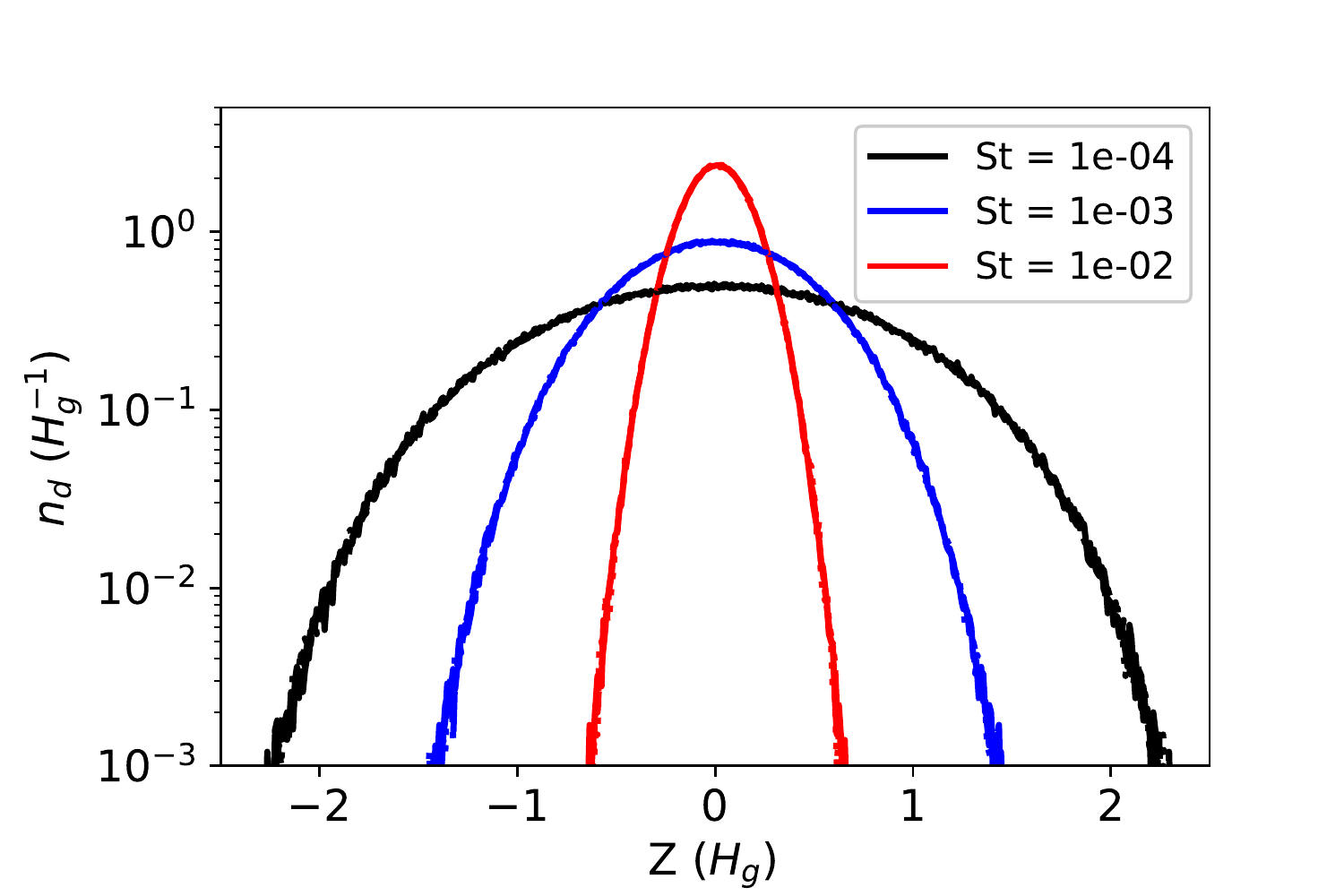}  
    \caption{Vertical dust density profile for different dust sizes for the antisymmetric gas flow profile. Results
    from three different times are shown, taken to be the same time as in Figure \ref{fig:scatter}, and they
    coincide almost exactly. The same applies to the case with symmetric gas flows. Density is normalized so that $\int n_d(z)dz=1$.}  \label{fig:scatter_z}
\end{figure}

\section{Simulation results}\label{sec:result}

We present the results form our Monte-Carlo simulation in this section. We start by describing the general properties of dust transport from representative cases in Section \ref{ssec:rep}, followed by characterizing all simulation results under the advection-diffusion framework in Section \ref{ssec:effective}.

\subsection{Representative Cases}\label{ssec:rep}

\begin{figure*}
    \centering
    \includegraphics[height=15cm, width=15cm]{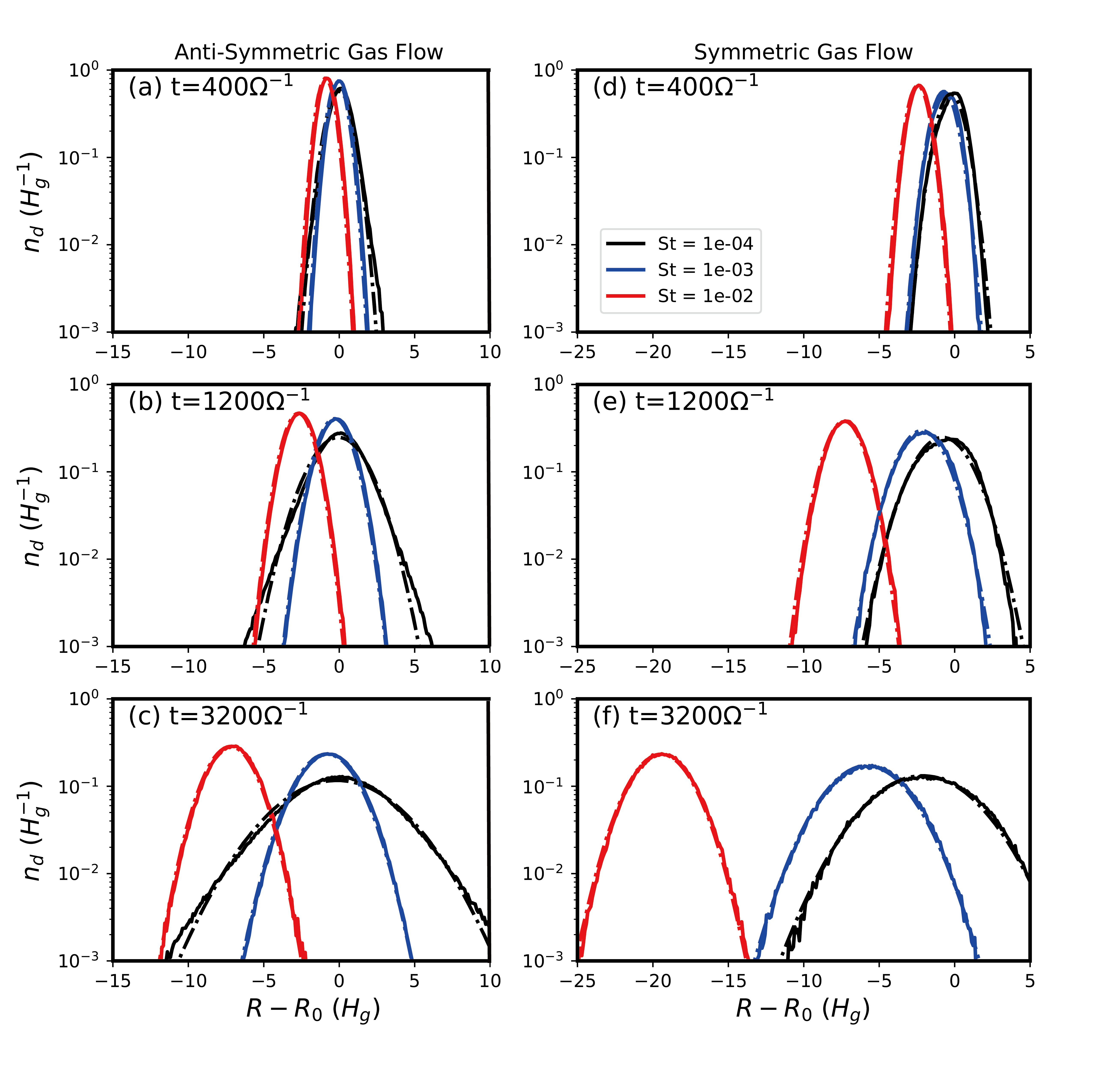}  
    \caption{Vertically-integrated radial distribution of dust at three different times
    ($t=400$, $1200$ and $3200\Omega^{-1}$), for different particle Stokes
    numbers ${\rm St}$ shown in Figure \ref{fig:scatter}. Measured results are shown in solid lines, whereas dash-dotted
    lines correspond to fitting results. Density is normalized so that $\int n_d(z)dz=1$.
    }  \label{fig:scatter_r}
\end{figure*}

In Figure \ref{fig:scatter}, we show
snapshots of particle positions for particles of different sizes at different times under anti-symmetric and symmetric gas flows. For illustrative purposes, we choose $\alpha=3\times10^{-4}$ and consider dust sizes with ${\rm St}=10^{-2}$, $10^{-3}$ and $10^{-4}$.
First of all, for both cases, we see that while the radial distribution of dust gets stretched to different extents at different heights,
affected by background gas flows, the vertical dust distribution remains stable over time. We further show and confirm in Figure
\ref{fig:scatter_z} that the vertical dust profile at different times almost exactly overlap. This indicates that the radial gas flow does
not affect the balance in the z-direction, and is the basis for us to formulate the theory of dust transport in the next section. This fact also arises from our local approximation, and is valid as long as we are only concerned with local
properties of dust transport.

In the anti-symmetric case, we clearly see that dust at upper and lower sides of the disk are pulled by background gas
flow towards opposite directions. This effect is more pronounced for smaller particles (${\rm St}=10^{-4}$), distorting the shape of the
particle cloud. Such distortion at different disk heights thus act effectively to disperse particles along radial directions. As particles
also oscillate and diffuse in the vertical direction, they spend finite time on each side of the disk. Therefore, as time progresses,
the overall shape of the particle cloud becomes more round due to vertical mixing, more similar to the case of pure diffusion.

In the symmetric case, there is a strong accretion flow at the midplane region, which gives rise to faster radial
drift for particles of all sizes. On the other hand, as particles diffuse to upper and lower parts of the disk, they are carried outward by the
gas flow. The combination gives a different shape of the particle cloud compared to the anti-symmetric case, but the overall
effect from the gas flow is similar: it acts as extra diffusive transport.

In many occasions, we are interested in the overall radial distribution of dust without caring about the detailed vertical structure.
This is particularly true when we would like to design global models of dust transport. We ask whether the effect from
the background gas flow can be incorporated into standard advection-diffusion framework by adding additional contributions to
dust drift and diffusion coefficients. In Figure \ref{fig:scatter_r}, we show the vertically-integrated radial profiles of dust particles
of various sizes and how they evolve with time, using the same data presented in Figure \ref{fig:scatter}. We fit particle radial
displacement $\langle [R(t)-R_0]^2\rangle$ as a function of time according to Equation (\ref{eq:r2}) to obtain $\bar{v}_{R}$ and
$\alpha_R$ for each dust species. In the meantime, we also show the expected Gaussian distribution profile from the fitting results,
namely $n_d(x)\propto\exp[-(x-v^{\textit{eff}}_{d,R}t)^2/2D^{\textit{eff}}_{d,R}t]$, in Figure \ref{fig:scatter_r} in dashed lines. We see that the measured dust density
distribution closely matches the expected Gaussian profile, thus demonstrating that the effect of background gas flow can indeed
be accurately described by the standard advection-diffusion framework.

\subsection[]{Effective Descriptions}\label{ssec:effective}

\begin{figure*}
    \centering
    \includegraphics[height=10.3cm, width=17cm]{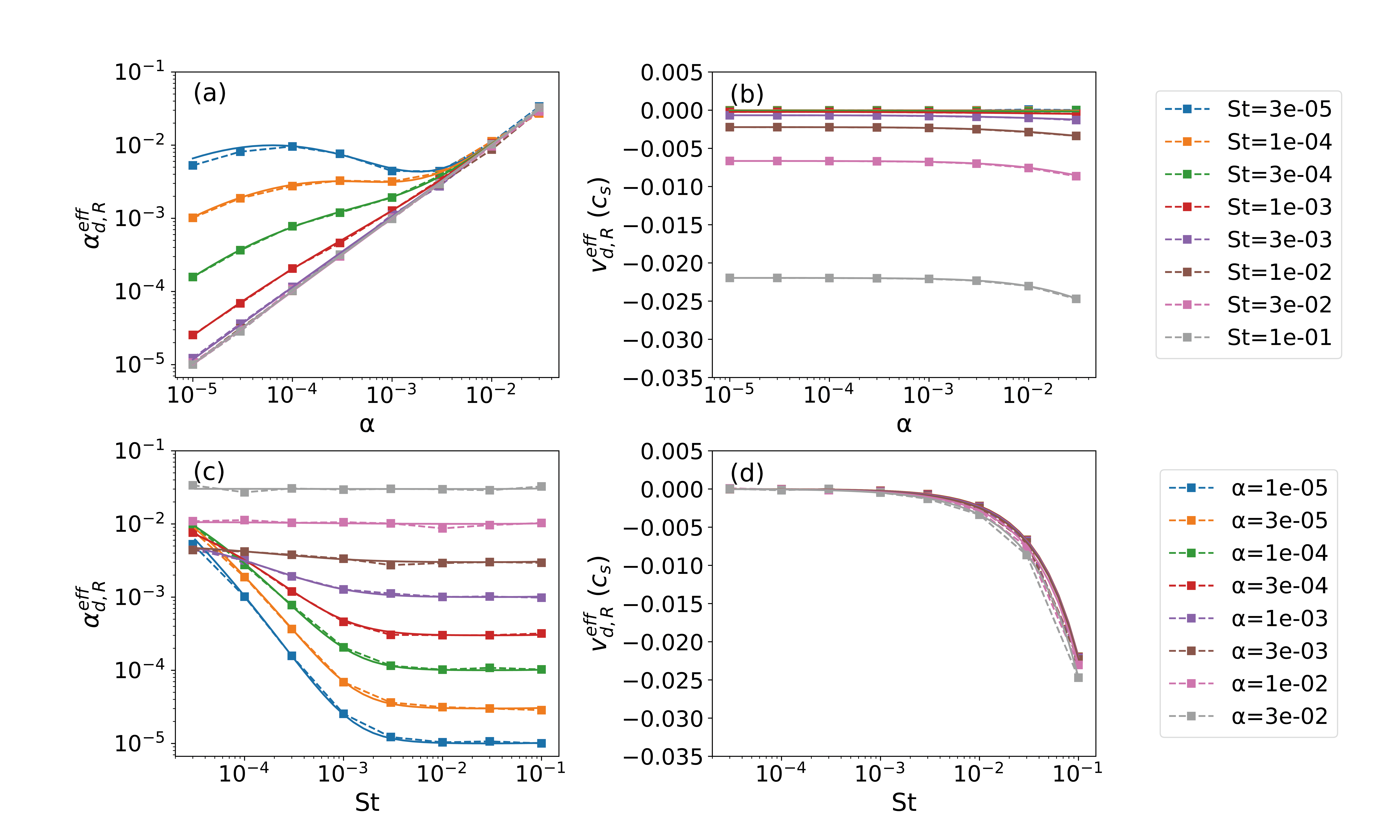}  
    \caption{Shearing sheet simulation results for dust diffusion in the inner region of PPD.
    (a) Relationship between effective radial {(pseudo-)}diffusion coefficient and $\alpha$ for different dust sizes. (b) Relationship between radial drift velocities and $\alpha$ for different dust sizes. (c) Relationship between dust effective radial {(pseudo-)}diffusion coefficient and dust size at different $\alpha$. (d) Relationship between dust radial drift velocities and dust size at different $\alpha$. In each panel, symbols connected by dashed lines represent simulation results, while solid lines represent results from theoretical calculations (see Section \ref{sec:calculation}).}\label{fig:r10_diff}
\end{figure*}

\begin{figure*}
    \centering
    \includegraphics[height=10.3cm, width=17cm]{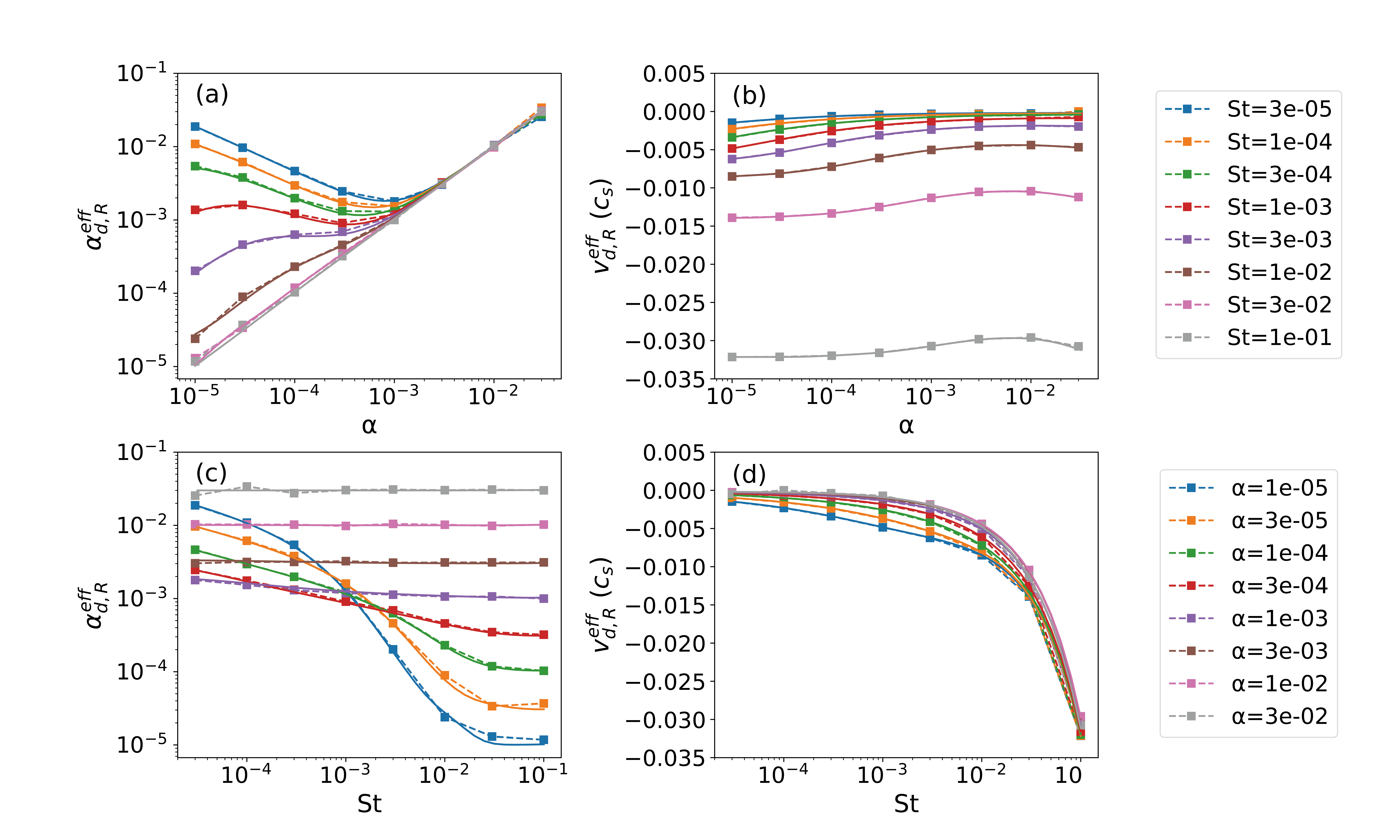}   
    \caption{Same as Figure \ref{fig:r10_diff} but for the symmetric case (18 AU).}\label{fig:r18_diff}  
\end{figure*}

The fact that the effect of wind-driven gas flows can be incorporated in to the standard advection-diffusion framework paves the way
for developing an effective description of global dust transport. From all our simulations, we fit for a radial drift velocity $v^{\textit{eff}}_{d,R}$
and effective radial {(pseudo-)}diffusion coefficient $D^{\textit{eff}}_{d,R}\equiv\alpha^{\textit{eff}}_{d,R}c_sH_g$, and show the results in Figures \ref{fig:r10_diff} and
\ref{fig:r18_diff} for the anti-symmetric and symmetric cases, respectively.


In the first panels for both figures, we show the effective radial {(pseudo-)}diffusion coefficient $\alpha^{\textit{eff}}_{d,R}$ versus the imposed gas diffusion
coefficient $\alpha$. We see a clear trend that for strongly coupled particles (${\rm St}\lesssim10^{-3}$) and at small background
turbulence ($\alpha<10^{-3}$), radial diffusion is enhanced. The enhancement
is more substantial towards smaller $St$, to reach an effective $\alpha^{\textit{eff}}_{d,R}\sim10^{-2}$, even when background turbulence
is as weak as $\alpha\sim10^{-5}$! This is the case because for such particles, they can reach a vertical height of $\gtrsim H_g$
and hence maximizing $D_R$ by experiencing the full range of inward and outward background gas flows. On the other hand,
for particles with larger Stokes number (${\rm St}\gtrsim10^{-3}$ in the anti-symmetric case and ${\rm St}\gtrsim10^{-2}$ in the symmetric
case), $\alpha^{\textit{eff}}_{d,R}\approx\alpha$. This is easily understood because these particles are largely confined to the midplane region,
and hence do not experience the full velocity range of radial gas flows. While a larger $\alpha$ can stir up these particles, it is also
strong enough to overwhelm the additional diffusion caused by radial gas flows. 

The discussion above can be equivalently viewed in the 3rd panels of the two figures. For background turbulence level
$\alpha\gtrsim10^{-3}$, the radial diffusion is largely governed by turbulence and is hardly affected by background gas flows.
For weaker turbulence, radial diffusion is more strongly enhanced for more strongly coupled particles. The level of enhancement
depends on the background gas flow profile, but eventually reaches $\alpha^{\textit{eff}}_{d,R}\sim10^{-2}$ in both anti-symmetric and symmetric cases.

It is also worth pointing out that, for very small dusts, their radial diffusion rate is always greater than some specific value. 
For instance, for the two gas flow velocity profiles we simulated, the $\alpha^{\textit{eff}}_{d,R}$ of dusts with ${\rm St} < 10^{-4}$ is always
greater than $10^{-3}$. This result can be proved analytically, as described in Section \ref{ssec:4.3}.

The second and fourth panel of the two figures show the influence of background gas flow on the mean radial drift velocity. We see
that overall, the influence on radial drift velocity is minor, as variation in $\alpha$ does not strongly change the mean radial drift speed.
This is especially true in the anti-symmetric case, as particles have equal probability to stay above and below the midplane, and hence
equally likely to be advected radially inward and outward by the background gas flow. On the other hand, for the symmetric case,
we see that radial drift is modestly enhanced for particles with ${\rm St}\lesssim0.01$ with weak background turbulence
$\alpha<10^{-3}$. This is specifically due to the midplane accretion flow with a velocity of $\sim3\times10^{-3}c_s$, and that
particles spend substantial fraction of time around the miplane region under weak turbulence. Particles with larger Stokes number are not strongly affected because their radial drift velocity from background radial pressure gradient is already larger than the velocity of the radial gas flows.

\section{Theoretical Calculations}\label{sec:calculation}

To develop further physical intuition and quantitative understanding about our simulation results, we first present a toy model in Section \ref{ssec:toy}, followed by a full calculation for any general gas flow profiles in Section \ref{ssec:full}.

\subsection[]{Toy model}\label{ssec:toy}

In our toy model, we simplify the vertical variation of dust radial drift velocity into a step function, set to be $v_1$ and $v_2$ above and below a certain height $z_0$, respectively. To isolate the role of this radial gas flows, we only consider turbulence in the vertical direction, which is essential for the discussion here, but ignore turbulence in the radial direction.

We denote the mean fractional time that a dust particle spends at locations above $z_0$ to be $X_+$, while the mean fractional time located below $z_0$ is denoted as $X_-$. In equilibrium state (balancing dust settling and diffusion in vertical direction), this fraction must be proportional to the total dust mass in these two layers.
For future convenience, we define
\bgeq\label{eq:chi}
\chi_0(z)\equiv\frac{\rho_d(z)}{\Sigma_d}\ ,
\edeq
as the normalized equilibrium dust density profile, where $\Sigma_d=\int_{-\infty}^\infty\rho_d(z)dz$ is the dust surface density. Note that $\chi_0(z)$ has a dimension of inverse of length so that $\int \chi_0(z)dz=1$. With this definition, we have
\bgeq\label{eq:defXpm}
X_+(z_0)=\int_{z_0}^\infty\chi_0(z)dz\ ,\quad X_-(z_0)=\int^{z_0}_{-\infty}\chi_0(z)dz\ .
\edeq
It is also straightforward to see that the mean radial drift velocity is
\bgeq\label{eq:vrtoy}
    v^{\textit{eff}}_{d,R} 
    = v_1X_+(z_0)+v_2X_-(z_0)\ .
\edeq

The effectiveness of radial diffusion is largely determined by the mean time a particle traverse the two layers above and below $z_0$ and returns. We may define this time as $t_{\rm cycle}$. Effectively, the process can be considered as a 1D random walk, and $t_{\rm cycle}$ corresponds to the mean time a particle cycles through two opposite directions. If this time is short, the particle only travels short distances radially as it undergoes vertical oscillations, leading to ineffective radial {pseudo-}diffusion. On the other hand, if the particle can stay on each side for long time, radial diffusion would be substantially enhanced due to its long effective radial mean free path.

Now we estimate $t_{\rm cycle}$. This time should depend on the position $z_0$, and a full cycle requires that on average, all particle have traversed both layers. At $z=z_0$, with dust density $\rho_{d,z}(z_0)$, we anticipate that over a time $\Delta t$, amount of mass through this location to be $\Delta\Sigma_d\approx\sqrt{D_{d,z}\Delta t}\rho_d(z_0)$. To allow all dust particles to traverse this location, $t_{\rm cycle}$ should satisfy 
\bgeq
\sqrt{D_{d,z}t_{\rm cycle}}\rho_d(z_0)\approx\Sigma_d\ .
\edeq
Using the definition (\ref{eq:chi}), we have $t_{\rm cycle}=[D_{d,z}\chi^2(z_0)]^{-1}$. We emphasize that this is only a rough estimate, and we defer to the next subsection for more rigorous calculations.

Over time $t\gg t_{\rm cycle}$, dust should have traversed the location $z_0$ over $n=t/t_{\rm cycle}$ times, spending an average amount of time $X_+t_{\rm cycle}$ above $z_0$ and $X_-t_{\rm cycle}$ below $z_0$ in each cycle. The variance of the radial position can be calculated to be
\bgeq\label{eq:r^2}
\begin{aligned}
    \langle R^2(t)\rangle - \langle R(t)\rangle^2 \approx &n\left[(v_1-v^{\textit{eff}}_{d,R})\times X_+t_{\rm cycle}\right]^2\\
    +&n\left[(v_2-v^{\textit{eff}}_{d,R})\times X_-t_{\rm cycle}\right]^2=2\Delta D^{\textit{eff}}_{d,R}t.
\end{aligned}
\edeq 
From the above, $\Delta D^{\textit{eff}}_{d,R}$ is found to be
\bgeq
\Delta D^{\textit{eff}}_{d,R}\approx\frac{1}{2}[(v_1-v^{\textit{eff}}_{d,R})^2X_+^2+(v_2-v^{\textit{eff}}_{d,R})^2X_-^2]t_{\rm cycle}\ ,
\edeq
which, after some algebra using Equation (\ref{eq:vrtoy}), can further be expressed as
\bgeq\label{eq:Drtoy}
\Delta D^{\textit{eff}}_{d,R}\approx\frac{[(v_1-v^{\textit{eff}}_{d,R})^2X_+^2+(v_2-v^{\textit{eff}}_{d,R})^2X_-^2]}{2D_{d,z}\chi^2(z_0)}
=\frac{(v_1-v_2)^2X_+^2X_-^2}{D_{d,z}\chi^2(z_0)}\ .
\edeq

There are several interesting features from the expression (\ref{eq:Drtoy}). First, unsurprisingly, $\Delta D^{\textit{eff}}_{d,R}$ scales as $(v_1-v_2)^2$, thus vertical variations in radial velocity strongly enhances radial diffusion.
Second, the effective radial diffusion coefficient from radial gas flows is {\it inversely proportional} to the turbulent diffusion coefficient $D_{d,z}$. This enter mainly through $t_{\rm cycle}\propto1/D_{d,z}$, and explains why the most enhanced radial diffusion is found only under weak turbulence. Third, under our toy model (with a step function variation in radial velocity), $\Delta D^{\textit{eff}}_{d,R}$ has sensitive dependence on the location $z_0$. It is strongly coupled with the dust density profile, which itself depends on $D_{d,z}$. Such inter-dependence also implies diverse and complex outcomes of dust transport in more realistic conditions, and calls for a more comprehensive theory as we develop below.

\subsection{General theory}
\label{ssec:full}

From the insight gained from the toy model, in this subsection we develop a general theory applicable to arbitrary radial flow profiles with more rigorous derivations. 

{We again start by assuming there is no turbulence in the radial direction ($D_{d,R}=0$).}
Consider the trajectory of a dust particle over time $t$, where $t$ is much larger than the analog of $t_{\rm cycle}$ in the previous subsection, which we call the dust vertical relaxation time. We divide $t$ by a large number of $N$ time intervals $\Delta t=t/N$ and denote the vertical position of the particle at each time as $z_1,z_2,\cdots,z_N$. Thus, the total distance it travels radially is
\begin{equation}\label{deltar}
    \Delta R = \sum_{i=1}^Nv_{d,R}(z_i)\Delta t.
\end{equation}
Taking ensemble average $\langle R^2(t)\rangle$, we obtain
\bgeq
\langle \Delta R^2(t)\rangle=\sum_{i=1}^N\sum_{j=1}^N\overline{v_{d,R}(z_i)v_{d,R}(z_j)}\Delta t^2\ .
\edeq
Notice that $\overline{v_{d,R}(z_i)v_{d,R}(z_j)}$ is the correlation function between $v_{d,R}(z_i)$ and $v_{d,R}(z_j)$, and also the result of this function should be symmetric about $z_i$ and $z_j$. 
Taking $\Delta t\rightarrow 0$, the above equation can be expressed into the following integral form
\begin{equation}
\begin{split}\label{eq:dR2}
\langle \Delta R^2(t)\rangle=&2\sum_{i=1}^N\sum_{j=i}^N\overline{v_{d,R}(z_i)v_{d,R}(z_j)}\Delta t^2\\
        =&2\int_0^t\text{d}t_1\int_{0}^{t-t_1}\text{d}t_2\overline{v_{d,R}(z(t_2))v_{d,R}(z(t_1+t_2))}\\
    =&2\int_0^t\text{d}t_1\int_{0}^{t-t_1}\text{d}t_2\overline{v_{d,R}(z(0))v_{d,R}(z(t_1))}\\
    =&2\int_0^t\text{d}t_1(t-t_1)\overline{v_{d,R}(z(0))v_{d,R}(z(t_1))}\ ,
\end{split}
\end{equation}
where we have used the time translational symmetry in the last two lines.

To proceed, we further define $\chi(z,z_1,t)$ as the normalized (probability) density of dust which resides at $z$ at time $t$, under the condition that it is located at $z_1$ at time zero. It can be normalized as $\int\chi(z,z_1,t)dz=1$. By definition, we have $\chi(z,z_1,0)=\delta(z-z_1)$. Moreover, over a period longer than the dust vertical relaxation time, dust should lose memory about its initial position, and hence we expect asymptotically, $\chi(z,z_1,t\rightarrow\infty)\approx\chi_0(z)$.
With this definition, we have
\begin{equation}
    \begin{aligned}
        \overline{\Delta R^2(t)}
        =&2\int_0^t\text{d}t_1(t-t_1)\int_{-\infty}^\infty\text{d}z_1\chi_{0}(z_1)\\
        &\times\int_{-\infty}^\infty dz_2\chi(z_2,z_1,t_1)v_{d,R}(z_1)v_{d,R}(z_2)\ .
    \end{aligned}
\end{equation}
To handle the integral, we can further define
\bgeq
F(z,z_1,T)\equiv \int_0^T\chi_{0}(z_1)\chi(z,z_1,t)\text{d}t-\chi_{0}(z)\chi_{0}(z_1)T\ ,
\edeq
\bgeq
H(z,z_1,T)\equiv \int_0^T t\chi_{0}(z_1)\chi(z,z_1,t)\text{d}t-\chi_{0}(z_1)\chi_{0}(z)T^2/2\ ,
\edeq
so that
\begin{equation}
    \begin{split}
        \overline{\Delta R^2(T)}
        =2T&\int_{-\infty}^\infty\text{d}z_1\int_{-\infty}^\infty dz_2F(z_1,z_2,T)v_{d,R}(z_1)v_{d,R}(z_2)\\
        -2\int_{-\infty}^\infty&\text{d}z_1\int_{-\infty}^\infty dz_2H(z_1,z_2,T)v_{d,R}(z_1)v_{d,R}(z_2)\\
        +T^2\int_{-\infty}^\infty&\text{d}z_1\int_{-\infty}^\infty dz_2\chi_0(z_1)\chi_0(z_2)v_{d,R}(z_1)v_{d,R}(z_2)\ .\\
    \end{split}
\end{equation}
Using the asymptotic properties of $\chi(z,z_1,T)$ for $T\rightarrow\infty$, we obtain
\bgeq
\frac{d F(z,z_1,T)}{dT}=\chi_{0}(z_1)\chi(z,z_1,T)-\chi_{0}(z)\chi_{0}(z_1)\rightarrow 0\ ,
\edeq
\bgeq
\frac{d H(z,z_1,T)}{dT}=T\chi_{0}(z_1)(\chi(z,z_1,T)-\chi_{0}(z))\rightarrow 0\ .
\edeq
Therefore, for large $T$, we expect
\bgeq
\begin{split}
F(z,z_1,T)&\approx F(z,z_1,\infty)\equiv F(z,z_1)\ ,\\
H(z,z_1,T)&\approx H(z,z_1,\infty)\equiv H(z,z_1)\ .
\end{split}
\edeq
This suggests that for large $T$,
\begin{equation}
    \begin{split}
        \overline{\Delta R^2(T)}\approx
        &2T\int_{-\infty}^\infty\text{d}z_1\int_{-\infty}^\infty dz_2F(z_1,z_2)v_{d,R}(z_1)v_{d,R}(z_2)\\
        &+T^2\bigg[\int_{-\infty}^\infty\text{d}z\chi_0(z)v_{d,R}(z)\bigg]^2+{\rm const}\ .\\
    \end{split}
\end{equation}
Comparing with $\overline{\Delta R^2}\rightarrow \bar{v}_R^2T^2+2\Delta D_RT$, it can be seen that the mean radial drift velocity is
\begin{equation}\label{eq:vr}
    v^{\textit{eff}}_{d,R}=\int\chi_0(z)v_{d,R}(z)dz\ ,
\end{equation}
which directly generalizes Equation (\ref{eq:vrtoy}), and the 
additional contribution from radial {pseudo-}diffusion is
\begin{equation}\label{eq:DeltaD0}
    \Delta D^{\textit{eff}}_{d,R}=\iint F(z_1,z_2)v_{d,R}(z_1)v_{d,R}(z_2) dz_1dz_2\ .
\end{equation}
The remaining task is to evaluate $\Delta D^{\textit{eff}}_{d,R}$.

\begin{figure*}
    \centering
    \includegraphics[height=7cm, width=15cm]{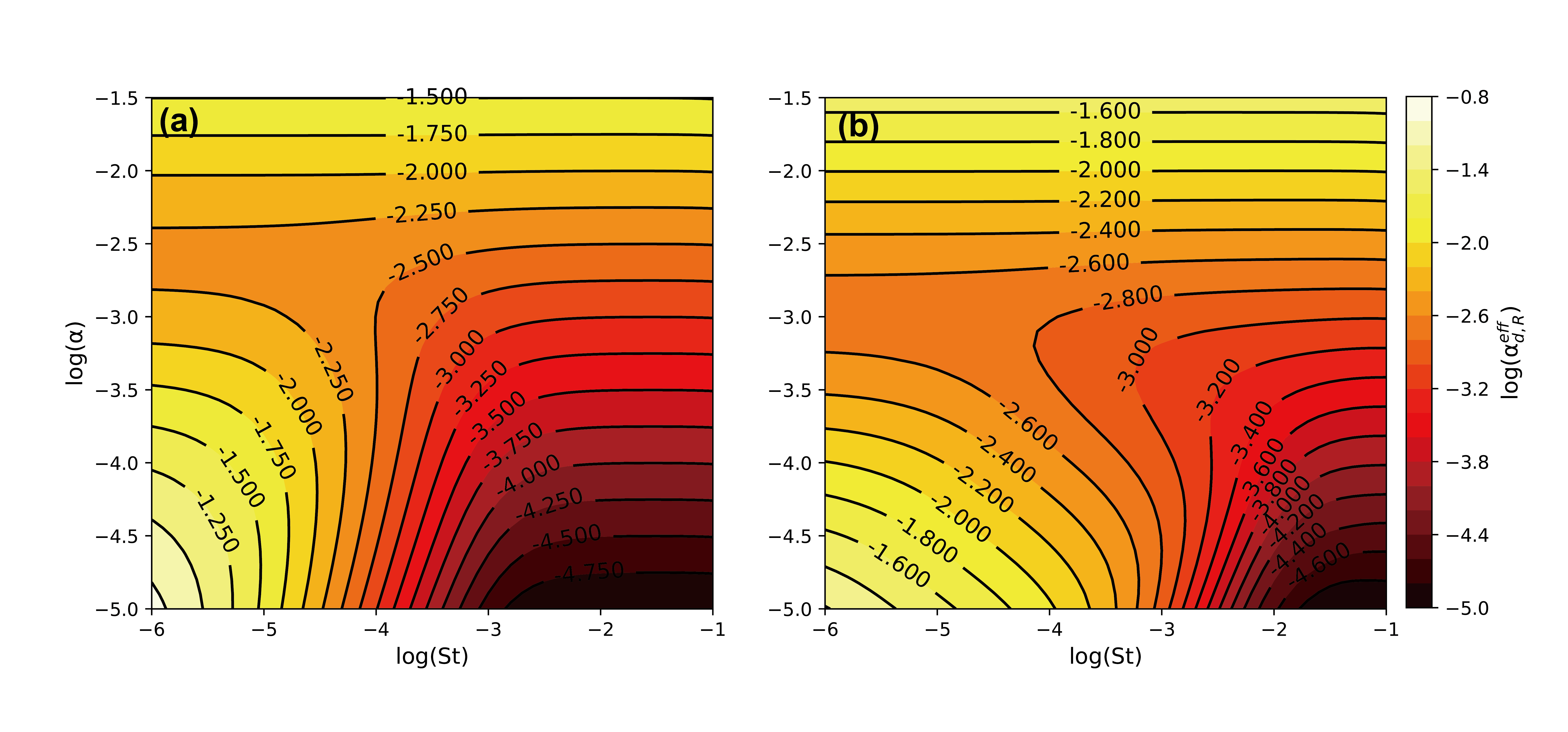}  
    \caption{Contour plots of the variation of $\alpha^{\textit{eff}}_{d,R}=D^{\textit{eff}}_{d,R}/H_gc_s$ with ${\rm St}$ and $\alpha$. The gas velocity distributions in (a) and (b) are the anti-symmetric and symmetric gas flow distribution cases in Figure \ref{fig:gasflow}, respectively.} \label{fig:contour}
\end{figure*}

Applying the vertical dust diffusion equation to $\chi(z,z_1,t)$ at given $z_1$,
and absorbing the correction from concentration diffusion to $v'_{d,z}$ from
Equation (\ref{eq:vd}), we have
\bgeq\label{eq:dustdif}
    \partial_t\chi=-\partial_z(\chi v'_{d,z}-D_{d,z}\partial_z\chi)\ .
\edeq
Multiplying both sides by $\chi_0(z_1)$ and integrating this equation over $t$ to some
large $T$, we obtain
\begin{equation}
    \begin{split}
        \chi_{0}(z_1)&[\chi_{0}(z)-\delta(z-z_1)]\\
        &=-\partial_z\left(F(z,z_1)v'_{d,z}(z)-D_{d,z}\partial_zF(z,z_1)\right),
    \end{split}
\end{equation}
where we have used $\chi(z,z_1,0)=\delta(z-z_0)$ and
that the equilibrium dust distribution $\chi_0(z)$ satisfies $\pa_z[\chi_0(z)v'_{d,z}]=D_d\pa_z\chi_0(z)$.
Let us further define
\bgeq
F(z,z_1)\equiv\chi_{0}(z_1)\chi_{0}(z)G(z,z_1)\ ,
\edeq
the above equation can be simplified to
\begin{equation}
    x_{0}(z)-\delta(z-z_1)=\partial_z\left[D_{d,z}x_{0}(z)\partial_zG(z,z_1)\right]\ .
\end{equation}
We integrate this equation over $z$ to some $z'$, resulting in
\bgeq
    D_{d,z}\chi_0(z)\partial_zG(z,z_1)=X_-(z)-\theta(z-z_1)\ ,
\edeq
where we have changed the variable back from $z'$ to $z$, and $\theta(z)$ is the step function. Therefore,
\bgeq
\begin{split}
    G(z,z_1)&=\int_{-\infty}^z\frac{X_-(z')-\theta(z'-z_1)}{D_{d,z}\chi_0(z')}dz'+C(z_1)\\
    &=\int_{-\infty}^\infty\frac{(X_-(z')-\theta(z'-z_1))(1-\theta(z'-z))}{D_{d,z}chi_0(z')}dz'+C(z_1)\\
    &=\int_{-\infty}^\infty\frac{(X_-(z')-\theta(z'-z_1))(X_-(z')-\theta(z'-z))}{D_{d,z}\chi_0(z')}dz'\\
    &\quad+\int_{-\infty}^\infty\frac{(X_-(z')-\theta(z'-z_1))(1-X_-(z'))}{D_{d,z}\chi_0(z')}dz'+C(z_1),
\end{split}
\edeq
where where $X_-(z)=\int_{-\infty}^z\chi_{0}(z')dz'$ as defined in (\ref{eq:defXpm}), and $C(z_1)$ is some function of $z_1$.
Note that $G(z,z_1)$ is symmetric with respect to $z$ and $z_1$, but the last two terms are only functions of $z_1$, this suggests that the sum of the last two terms must be a constant. This constant must be zero, which would otherwise violate the translational symmetry for $\Delta D^{\textit{eff}}_{d,R}$ from Equation (\ref{eq:DeltaD0}). In other words, $\Delta D^{\textit{eff}}_{d,R}$ should remain unchanged when $v_{d,R}(z)\rightarrow v_{d,R}(z)+{\rm const}$. Therefore, we arrive at
\begin{equation}
    \begin{aligned}
    &G(z,z_1)=\int_{-\infty}^\infty\frac{(X_-(z')-\theta(z'-z))(X_-(z')-\theta(z'-z_1))}{\chi_0(z')D_{d,z}}dz'.
    \end{aligned}
\end{equation}

\begin{figure*}
    \centering
    \includegraphics[height=14.5cm, width=16cm]{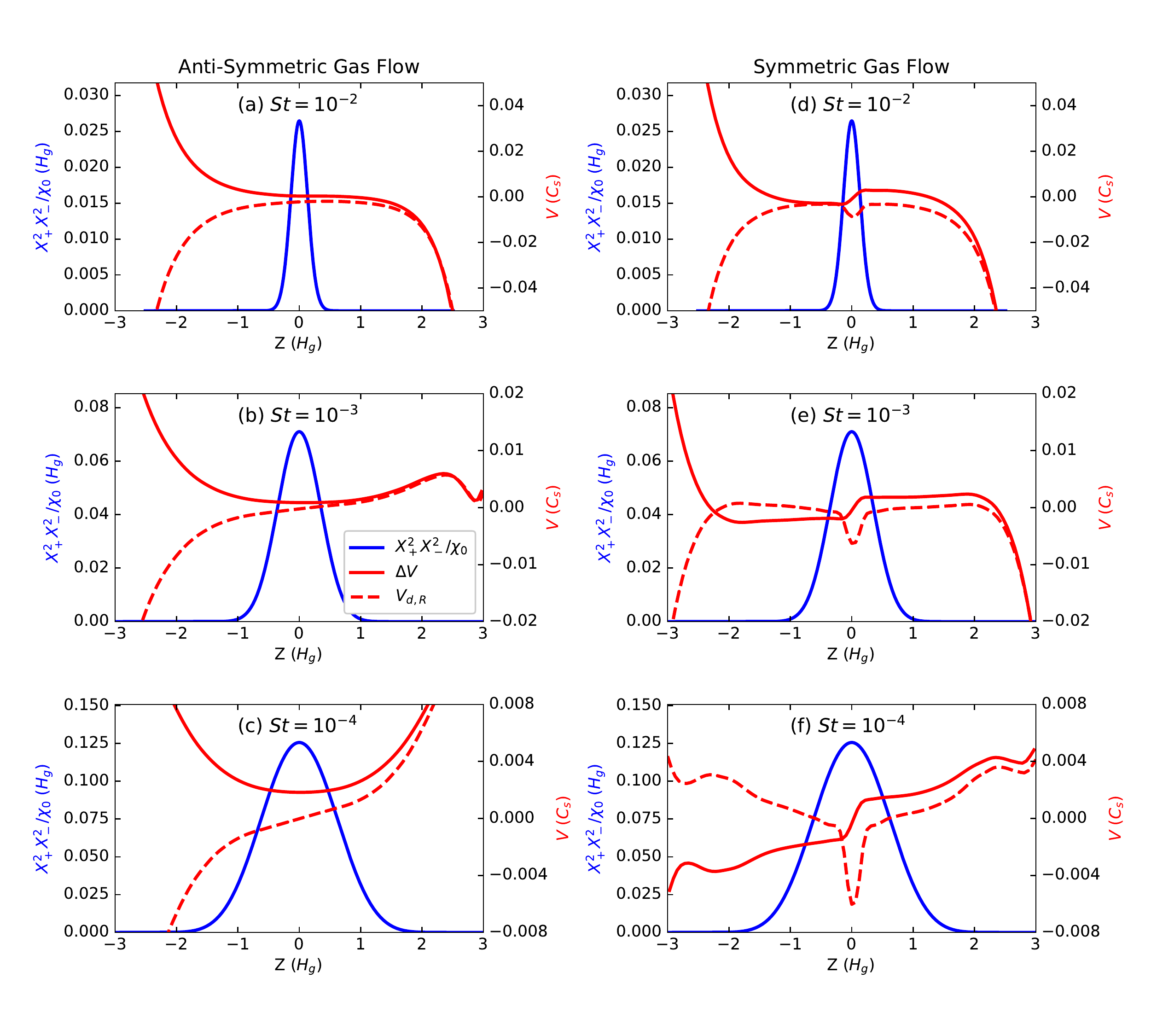}  
    \caption{Vertical profiles of major factors in Equation (\ref{eq:DeltaD}) contributing to the effective dust radial {(pseudo-)}diffusion coefficient $\Delta D_{R}$ for the anti-symmetric (left) and symmetric (right) cases. Gas turbulence is fixed to $\alpha=3\times10^{-4}$, and three rows correspond to different dust sizes. Blue solid lines show $X_+^2X_-^2/\chi_0$ factor, red solid lines show the $\Delta V$ factor, while red dashed lines indicate mean dust drift speed at each height.} \label{fig:vprof}
\end{figure*}

Substituting the above to Equation (\ref{eq:DeltaD0}), the integral on $z_1$ and $z_2$ are readily separable and are identical, we find after some algebra the final expression for $\Delta D^{\textit{eff}}_{d,R}$ as
\begin{equation}\label{eq:DeltaD}
    \Delta D^{\textit{eff}}_{d,R} = \int \frac{X_+^2X_-^2\Delta V(z)^2}{\chi_0(z)D_{d,z}(z)}dz\ ,
\end{equation}
where $X_\pm(z)$ are defined in (\ref{eq:defXpm}), and
\bgeq
    \Delta V(z) \equiv \bar{V}_+(z) - \bar{V}_-(z),
\edeq
where $\bar{V}_\pm(z)$ are defined as follows
\bgeq
\begin{aligned}
    &\bar{V}_+(z) = \int_z^\infty \chi_0(z)v_{d,R}(z)dz/X_+(z)\ ,\\
    &\bar{V}_-(z) = \int_{-\infty}^z\chi_0(z)v_{d,R}(z)dz/X_-(z)\ .
\end{aligned}
\edeq
Clearly, $V_\pm(z)$ represents density-weighted mean radial drift velocity above and below height $z$.
 We see that our final result (\ref{eq:DeltaD}) closely resembles the result from the toy model, retaining the $\Delta V^2/D_{d,z}$ scaling as discussed earlier. The integral over $z$ yields the overall weighting factor that applies to any radial flow profile. Also note that although we assumed constant $D_{d,z}$ in the simulations, our analytic calculation applies also to the general case with height-dependent $D_{d,z}(z)$. 
 
{Finally, we consider turbulence in the radial direction with non-zero $D_{g,R}$ (and hence non-zero $D_{d,R}$). Because its contribution to the calculation of $\langle\Delta R^2(t)\rangle$ in Equation (\ref{eq:dR2}) is additive at each height in the integral, it can be directly separated out, yielding a weighted average $\overline{D_{d,R}}$ given by Equation (\ref{eq:DdRbar}). Therefore,}
the full radial {pseudo-}diffusion coefficient is simply given by $D_{d,R}^{\textit{eff}}\equiv \overline{D_{d,R}}+\Delta D_{d,R}^{\textit{eff}}$, or Equation (\ref{eq:Dtot}).

\subsection[]{General properties and comparison with simulations}\label{ssec:4.3}

To verify our analytical results, in Figures \ref{fig:r10_diff} and \ref{fig:r18_diff}, we further show $v_{d,R}^{\textit{eff}}$ and $D_{d,R}^{\textit{eff}}$ from analytical calculations above. We see they are in almost perfect agreement with simulation results, thus precisely validating our theory. With our analytical tool, we further show in Figure \ref{fig:contour} contour plots of $\alpha_{d,R}^{\textit{eff}}=D_{d,R}^{\textit{eff}}/H_gc_s$ as a function of ${\rm St}$ and $\alpha$ for both anti-symmetric and symmetric cases, supplementing the results shown in Figures \ref{fig:r10_diff} and \ref{fig:r18_diff}. The value of $\alpha_{d,R}^{\textit{eff}}$ at ${\rm St}=0.1$ is largely dominated by background turbulence, and as we consider more tightly coupled particles with smaller ${\rm St}$, the contrast with the ${\rm St}=0.1$ result demonstrate the influence of radial gas flows. The anti-symmetric and symmetric cases show similar trend but differ in details, as discussed in Section \ref{ssec:effective}.

With our analytical tool, we can further quantify to what extent the given gas radial flow profile contributes to the effective {(pseudo-)}diffusion coefficient. From Equation (\ref{eq:DeltaD}), the contribution is primarily through the $[V_+(z)-V_-(z)]^2/D_d$ factor, weighted by $[X_+(z)X_-(z)]^2/\chi_0(z)$ factor. In Figure \ref{fig:vprof}, we plot the vertical profiles of these factors for both the antisymmetric and symmetric radial flow profiles, using representative values of $\alpha=3\times10^{-4}$ and ${\rm St}=10^{-2}, 10^{-3}, 10^{-4}$, same as in Figures \ref{fig:scatter} and \ref{fig:scatter_r}. 

The vertical profiles of the $[X_+(z)X_-(z)]^2/\chi_0(z)$ factor are relatively simple, showing smooth centrally-peaked Gaussian-like structure. The profile becomes wider for more tightly coupled particles, roughly reflecting the dust vertical density profile shown in Figure \ref{fig:scatter_z}. Regions towards larger $z$ are strongly suppressed as one takes the squares, and hence the profiles are narrower than those in Figure \ref{fig:scatter_z}.

The profiles for $v_{d,R}(z)$ and $\Delta V$ are more complex. For particles with small ${\rm St}$, the $v_{d,R}$ profiles largely coincides with the radial flow profile of background gas shown in Figure \ref{fig:gasflow}. Increasing dust sizes, radial drift due to background gas pressure gradient becomes more significant, and the total drift velocity profile can become asymmetric about the disk midplane. Eventually (${\rm St}\gtrsim0.01$), the profile becomes symmetric again as radial drift is completely dominated by background pressure gradient. The profiles of $\Delta V$ reflects a convolution of $v_{d,R}$ profile with density profile $\chi_0$ followed by differencing. Note that it is symmetric about midplane for an anti-symmetric $v_{d,R}$ profile and vice versa, and is generally more smooth near the midplane region due to the convolution.

Overall, we see that contribution to $\Delta D_{d,R}^{\textit{eff}}$ from the $\Delta V^2/D_{d,z}$ term is generally stronger towards larger heights, whereas the $[X_+(z)X_-(z)]^2/\chi_0(z)$ factor strongly suppress contributions from larger heights. The combination of these favors more vertical spread of the dust layer, and more significant radial velocity variation near the midplane region, to achieve larger $\Delta D_{d,R}^{\textit{eff}}$. This explains why the anti-symmetric case and symmetric case yield similar $\Delta D_{d,R}^{\textit{eff}}$ for ${\rm St}=10^{-4}$, while the symmetric case yields larger $\Delta D_{d,R}^{\textit{eff}}$ for ${\rm St}=10^{-3}$.

{We further note that the expression (\ref{eq:DeltaD}) for $\Delta D_{d,R}^{\textit{eff}}$ only depends on $D_{g,z}$ (and hence $D_{d,z}$) but not $D_{g,R}$ (and hence $D_{d,R}$). This means that our results can be easily applied to anisotropic diffusion, where $D_{g,z}\neq D_{g,R}$ (e.g., \citealp{Stoll_etal17}), as their contributions to the total effective (pseudo-)diffusion coefficients are separate and independent. In this case, one can reuse Figure \ref{fig:contour} by treating $\alpha=D_{g,z}$, while changing the additive contribution from $\overline{D_{d,R}}$.}

Since the most effective case that enhances radial {pseudo-}diffusion is with small dust, we here discuss a special regime applicable for tiny particles with ${\rm St}\ll\alpha$. This regime also applies to gas diffusion (e.g., of chemical species) itself. In this case, $\chi_0(z)\approx \rho_g(z)/\Sigma_g$ and $D_{d}\approx D_g$. If we assume that $D_g$ is constant in the vertical direction, then the total radial {pseudo-}diffusion coefficient is
\bgeq\label{eq:dr}
    D_{d,R}^{\textit{eff}} = D_g+\frac1{D_g}\int\frac{X_+^2X_-^2\Delta V(z)^2}{\chi_0(z)}dz\geq2\sqrt{\int\frac{X_+^2X_-^2\Delta V(z)^2}{\chi_0(z)}dz}\ .
\edeq
Note that here $\Delta V$ is entirely from the radial gas flow distribution, and hence this radial gas flow distribution completely determines the minimum radial {pseudo-}diffusion coefficient of small dusts. When applying such flow profiles for the anti-symmetric and symmetric cases, we find $D_{d,R}^{\textit{eff}}\geq 5\times 10^{-3}$ and $D_{d,R}^{\textit{eff}}\geq 2\times10^{-3}$ respectively. We can identify from our simulations and from Figure \ref{fig:contour} that these are all satisfied for $St\lesssim\alpha$. For diffusion of chemical species, it generally suffices to consider the values indicated in the left-most region of Figure \ref{fig:contour} for an estimate of the effective radial {(pseudo-)}diffusion coefficient.

Under the same assumption for small dust (${\rm St}\ll\alpha$), it is also interesting to ask under what vertical gas flow profile can one maximize $\Delta D_{d,R}^{\textit{eff}}$. This is a pure mathematical problem and is discussed in Appendix \ref{sec:appendix}, and it turns out given the level of vertical variations in the gas flow profiles, the shape of the gas radial flow profile we adopted is not far from optimal, and the resulting $D_{d,R}^{\textit{eff}}$ falls short of the maximum value possible by within a factor of 2.

\section[]{Discussion}\label{sec:discussion}

\subsection[]{Limitations and generalizations}

For the sake of simplicity, we have restricted ourselves to local models. This is partly because the vertical profile of gas flows varies with radius, as seen in Figure \ref{fig:gasflow}, and choosing certain specific radii helps us concentrate on the physics dust transport with better transparency. Our results, that the diffusive nature of vertically-integrated radial dust transport makes it straightforward to generalize to global disk models, provided that global gas flow structure and level of turbulence are known. 

Our simulations and calculations have neglected backreaction from dust to gas, and when taken into account, it will affect the gas flow profiles and hence global transport depending on the level of dust settling, dust abundance and size distributions (e.g., \citealp{Tanaka2005,BaiStone2010,Dipierro2018}). On the other hand, our analytical theory is applicable as long as gas flow profiles are known, which can incorporate the effect of dust backreaction as needed.

The wind-driven accretion/decretion profiles we consider in this work are representative, but not exclusive. For instance, amplification of horizontal magnetic field due to the HSI depends on initial conditions. In our choice from \citetalias{Bai_2017}, toroidal field only changes sign once in the disk, and hence the resulting flow structure (from vertical gradient of toroidal field) is relatively simple.
There can be situations where toroidal field changes sign multiple times (e.g., Appendix A in \citetalias{Bai_2017}, \citealt{bethune2017global}),
which could yield even more complex gas flow structure with multiple accretion and decretion streams over the vertical extent
of the disk, and sometimes the decretion flow may be located at the midplane. We anticipate even stronger enhancement of dust {pseudo-}diffusion in such cases, and potentially efficient outward dust transport.

\subsection[]{Applications}

Our simulation results and analytical model can be broadly applicable to many aspects of transport problems in PPDs, as we briefly discuss below.

{\it Models of grain growth and transport:}
Our results indicate that when modeling global dust transport, the radial {pseudo-}diffusion coefficient can be strongly size dependent, and is strongly enhanced for small dust. This effect can potentially be important for studying coupled grain growth when coupled with a dust transport model (e.g., \citealp{Garaud2007,Birnstiel2010,Misener2019}), and should be considered in general models dust transport for astrophysical and planetary science applications (e.g. \citealp{Gail2001,2002A&A...384.1107B,Cuzzi2003,YangCiesla12,Estrada2016}). 

{\it Constraining turbulent diffusion:} Our results suggest that the effective radial {(pseudo-)}diffusion coefficient and vertical diffusion coefficients for dust can differ significantly, especially for small dust particles. This is in line with the relatively high level of radial diffusivity inferred by modeling dust ring width \citep{Dullemond_2018,Rosotti_2020}, and the low-level of turbulence inferred from substantial dust settling (e.g., \citealp{Pinte_2015}), although the observations apply to larger dust particles, and additional physics may be needed (e.g., \citealt{baehr2021particle}).

{\it Large-scale mixing of solids:} We note that the enhanced radial {pseudo-}diffusion for strongly coupled dust applies for chondrule-sized particles in the inner solar system.
In the representative cases we consider, we do not observe significant outward dust transport from our simulations, thus these cases are insufficient to account for evidence of large-scale mixing in the solar system. However, as just discussed, the HSI-induced flow structure can be more complex (\citetalias{Bai_2017}, \citealt{bethune2017global}) to allow more efficient outward transport. Alternatively, outward transport can be realized in presence of ``coronal accretion", where accretion flow is concentrated in the surface, together with a decretion flow found in the midplane layer (e.g., \citealt{zhu2018global}).

{\it Transport of chemical species:} The limit of strongly coupled dust is equivalent to individual chemical species in the gas. Turbulent mixing is known to play an important role in disk chemistry, which can enhance the abundance of many gas and ice phase species \citep{Semenov2011,Furuya2014}, affect the evolution of solids near the snow lines (e.g., \citealp{Krijt2016,Krijt2018,Xu2017}) and hence potentially planetesimal formation \citep{2017A&A...608A..92D,Schoonenberg2018}, help explain the discrepancy of D/H ratio in solar system bodies \citep{Albertsson2014,Furuya2013}. Our results would suggest using highly different values of effective diffusion coefficients in vertical and radial directions for more realistic model calculations.

\section{Summary}\label{sec:summary}

In this paper, we have presented the results of Monte Carlo simulations of dust transport in the $r-z$ plane for dust with different sizes under two representative vertical profiles of gas accretion/decretion flow structures resulting from wind-driven accretion, together with external sources of turbulence. Our main findings are as follows.
\begin{itemize}
\item The vertical distribution of dust remains unaffected under additional accretion/decretion gas flow structures.
\item When integrated vertically, the effect of additional gas flow structures can be well incorporated into the standard advection-diffusion framework in the radial direction, with modified radial drift speeds and {(pseudo-)}diffusion coefficients. 
\item The accretion/decretion flow structure can significantly enhance the radial diffusion of small dust (${\rm St}\lesssim10^{-3}$). Even with weak turbulence $\alpha\lesssim10^{-4}$, the effective radial ({pseudo-})diffusion can reach up to $\alpha_{d,R}^{\textit{eff}}\sim10^{-2}$ in the representative examples we consider.
\item Radial drift of small dust (${\rm St}\lesssim10^{-2}$) is modestly enhanced in the presence of midplane accretion flow (the ``symmetric" case)
that is characteristic of wind-driven accretion in the outer disks.
\end{itemize}

We have also developed a general analytic theory that can be employed to calculate the mean dust radial drift speeds and effective dust radial {(pseudo-)}diffusion coefficients
for any vertical profiles of gas flow velocity and diffusion coefficient.
Key results are given in Equation (\ref{eq:vr}) for the radial drift speed, and Equation (\ref{eq:DeltaD}) for the {pseudo-}diffusion coefficient due to the presence of radial gas flows. For strongly coupled dust, we also provide a lower limit in total radial dust {pseudo-}diffusion coefficient, and discuss the optimal gas flow profiles to maximize radial diffusion.
The theory precisely reproduces our simulation results, and can be considered
as a general framework to study dust transport when considering more realistic gas flow structures in PPDs, with broad applications in studying the transport and evolution of solids and chemical species in PPDs.

Our study highlights the additional complication to dust transport brought by the accretion processes, especially associated with wind-driven angular momentum transport mediated by non-ideal MHD processes. It also calls for efforts to improve our understandings of disk microphysics, which ultimately determines the gas dynamics and flow structure in PPDs.

\section*{Acknowledgements}

We acknowledge the anonymous referee for a prompt report with helpful suggestions.
This work is supported by the National Key R\&D Program of China (No. 2019YFA0405100). 

\section*{Data Availability}
The data underlying this article are available in the article.
%and in its online supplementary material.



\bibliographystyle{mnras}
\bibliography{dustdiff} % if your bibtex file is called example.bib

%\begin{thebibliography}{99}
%\bibitem[\protect\citeauthoryear{Author}{2013}]{author2013}
%\bibitem[\protect\citeauthoryear{Jones}{2015}]{jones2015}
%Jones C.~D., 2015, Journal of Interesting Stuff, 17, 198
%\bibitem[\protect\citeauthoryear{Smith}{2014}]{smith2014}
%Smith A.~B., 2014, The Example Journal, 12, 345 (Paper I)
%\end{thebibliography}

%%%%%%%%%%%%%%%%%%%%%%%%%%%%%%%%%%%%%%%%%%%%%%%%%%

%%%%%%%%%%%%%%%%% APPENDICES %%%%%%%%%%%%%%%%%%%%%

\appendix

\section{Radial flow profile to maximize radial dust diffusion}\label{sec:appendix}

In this appendix, we consider the following mathematical problem. Under the following two constraints
\bgeq\label{eq:a1}
\int_{-\infty}^\infty \chi_0(z)v_{d,R}(z) dz = 0\ ,
\edeq
\bgeq\label{eq:a2}
\int_{-\infty}^\infty \chi_0(z) v_{d,R}^2(z) dz = \sigma^2\ ,
\edeq
where $\sigma$ is constant, what functional form of $v_{d,R}(z)$ could maximize
\bgeq
\Delta D_{d,R}^{\textit{eff}}=\int\frac{X_+^2X_-^2\Delta V(z)^2}{\chi_0(z)} dz\ .
\edeq
From Equation (\ref{eq:a1}), we have
\bgeq\label{eq:a4}
\begin{aligned}
X_+X_-\Delta V(z) &= \int_{-\infty}^\infty (X_-(z)-\theta(z-z'))\chi_0(z')v_{d,R}(z')dz'\\
&= -\int_{-\infty}^z\chi_0(z')v_{d,R}(z')dz'.
\end{aligned}
\edeq
By conducting a variation of $v_{d,R}$ (denoted as $\delta v$)
on $\int_{-\infty}^{\infty}\Delta V(z)^2/\chi_0(z)\text{d}z$, we obtain
\begin{equation}
    \begin{aligned}
    &\delta\int_{-\infty}^{\infty}\frac{X_+^2X_-^2\Delta V(z)^2}{\chi_0(z)}  dz\\
    &=-2\int_{-\infty}^{\infty}dz\frac{X_+^2X_-^2\Delta V(z)}{\chi_0(z)}\int_{-\infty}^{z}dz'\chi_0(z')\delta v(z')\\
    &=-2\int_{-\infty}^{\infty}d\left(\int_{-\infty}^zdz'\frac{X_+X_-\Delta V(z')}{\chi_0(z')}\right)\int_{-\infty}^{z}dz'\chi_0(z')\delta v(z')\\
    &=2\int_{-\infty}^{\infty}dz\chi_0(z)\delta v(z)\int_{-\infty}^{z}dz'\frac{X_+X_-\Delta V(z')}{\chi_0(z)}.
    \end{aligned}
\end{equation}
In order to obtain the extreme value,
\bgeq
\delta\int_{-\infty}^{\infty}\frac{X_+^2X_-^2\Delta V(z)^2}{\chi_0(z)}  dz = 0
\edeq
for any $\delta v(z)$ that maintains Equation (\ref{eq:a1}) and (\ref{eq:a2}), 
there should be constants $a$, $b$ such that the following equation holds
\bgeq
\int_{-\infty}^z \frac{X_+X_-\Delta V(z')}{\chi_0(z')}dz' = a + bv_{d,R}.
\edeq
From (\ref{eq:a4}), this is equivalent to
\bgeq\label{eq:a8}
    v_{d,R}(z)\chi_0(z)= -b\frac{d}{dz}\left(\chi_0(z)\frac{dv_{d,R}(z)}{d z}\right)\ .
\edeq
Using Equation (\ref{eq:rhod}),
\bgeq
\frac{d\chi_0(z)}{dz}=\left(-\frac{z}{H_g^2} + \frac{\bar{v}_{d,z}}{D_{d,z}}\right)\chi_0(z)=-\left(1+\frac{{\rm St}}{\alpha}\right)\frac{z}{H_g^2}\chi_0(z),
\edeq
Equation (\ref{eq:a8}) becomes
\bgeq\label{eq:a10}
\frac{d^2v_{d,R}(z)}{dz^2}-\left(1+\frac{{\rm St}}{\alpha}\right)\frac{z}{H_g^2}\frac{dv_{d,R}(z)}{dz}+\frac{v_{d,R}(z)}{b}=0.
\edeq
In the limit of strongly-couple dust (${\rm St}\ll\alpha$),
the coefficient $1+{\rm St}/\alpha\approx1$ is a constant.
In this case, the solution of Equation (\ref{eq:a10}) is 
\bgeq
v_{d,R}(z) \propto H_n(z/H_g)\ ,
\edeq
where 
\bgeq
H_n(\xi) = (-1)^ne^{\xi^2/2}\frac{{\rm d}^n e^{-\xi^2/2}}{{\rm d}\xi^n}
\edeq
is the $n$th Hermite polynomial with $n=1,2,3,\cdots$. The corresponding diffusion coefficient is
\bgeq
\Delta D_{d,R}^{\textit{eff}}=\int \frac{X_+^2X_-^2\Delta V(z)^2}{D_g\chi_0(z)}\text{d}z=\frac{\sigma^2H_g^2}
{nD_g}\ .\label{eq:Dmax}
\edeq
Therefore, the maximum radial {pseudo-}diffusion coefficient that an anti-symmetrically distributed gas flow can bring to the dust is achieved by choosing $n=1$, while the maximum radial {pseudo-}diffusion coefficient that a symmetrically distributed gas flow can bring to the dust is achieved for $n=2$.

In particular, the $\sigma^2$ of the gas flow distribution we use are $9.3\times 10^{-6} c_s^2$ for anti-symmetric case and $4.3\times10^{-6} c_s^2$ for symmetric case. Applying to Equation (\ref{eq:DeltaD}) for strongly coupled dust and compare to (\ref{eq:Dmax}), we find an effective $n=1.35$ for the anti-symmetric case and an effective $n=3.63$ for the symmetric case. This suggests that the profiles of radial gas flows we adopted are not far from being optimal.

%%%%%%%%%%%%%%%%%%%%%%%%%%%%%%%%%%%%%%%%%%%%%%%%%%

% Don't change these lines
\bsp	% typesetting comment
\label{lastpage}
\end{document}